\documentclass[aps,twocolumn,pre,preprintnumbers,amsmath,amssymb]{revtex4-2}
\usepackage[breaklinks=true,colorlinks=true
,urlcolor=blue
,anchorcolor=blue
,citecolor=blue
,filecolor=blue
,linkcolor=blue
,menucolor=blue
,pagecolor=blue
,linktocpage=true
,pdfproducer=medialab
,pdfa=true
]{hyperref}
\usepackage{bbold}
\usepackage{hyperref}
\usepackage[dvipdfmx]{graphicx}
\usepackage{amsmath}
\usepackage{graphicx,float,color}
\definecolor{c1}{RGB}{0,148,17}
\definecolor{c2}{RGB}{255,137,0}

\newcommand{\ket}[1]{| #1 \rangle}
\newcommand{\bra}[1]{\langle #1 |}

\newcommand{\be}{\begin{equation}}
\newcommand{\ba}{\begin{eqnarray}}
\newcommand{\ea}{\end{eqnarray}}
\newcommand{\ee}{\end{equation}}

\newcommand{\la}{\langle}
\newcommand{\lb}{\rangle}
\newcommand{\bea}{\begin{eqnarray}}
\newcommand{\eea}{\end{eqnarray}}
\newcommand{\bes}{\begin{equation*}}
\newcommand{\beas}{\begin{eqnarray*}}
\newcommand{\eeas}{\end{eqnarray*}}
\newcommand{\bas}{\begin{array*}}
\newcommand{\eas}{\end{array*}}
\newcommand{\ees}{\end{equation*}}

\newcommand{\bpm}{\begin{pmatrix}}
\newcommand{\epm}{\end{pmatrix}}
\newcommand{\bbm}{\begin{bmatrix}}
\newcommand{\ebm}{\end{bmatrix}}

\usepackage{bm}% bold math

\newcommand{\tr}{{\rm tr}}

\begin{document}

% \begin{flushleft}
% IPM/P-2024/xxx
% \end{flushleft}

\title{Information Scrambling in ‌Bosonic Gaussian Dynamics}
\author{Ali Mollabashi and Saleh Rahimi-Keshari}

\affiliation{
School of Quantum Physics and Matter, Institute for Research in Fundamental Sciences (IPM),
1939555311, Tehran, Iran
}

\begin{abstract}
We investigate the dynamics of information scrambling in bosonic systems undergoing Gaussian unitary evolution associated with quadratic Hamiltonians. For initial Gaussian states, we observe the disappearance of the memory effect in the entanglement dynamics of disjoint blocks under Gaussian random local dynamics. In addition, we show that randomness in the Hamiltonian causes the tripartite mutual information to saturate at relatively large negative values. Therefore, despite being integrable, these systems exhibit information-scrambling diagnostics that mirror those observed in chaotic systems. We note, however, that random quadratic Hamiltonians can have a component exhibiting Wigner-Dyson energy-level statistics; for non-Gaussian states within the corresponding subspace, these systems can display chaotic behavior. Our results provide insight into the Gaussian dynamics of continuous-variable systems, which are useful and available resources for quantum information processing.

\end{abstract}
\maketitle
%%%%%%%%%%%%%%%%%%%%%%%%%%%%%%%%%%%%%%%%%%%%%%%%%%%%%%%%%%%%%%%%%%%%%%%%%%%%%%%%%%%%%%%%%%%%%%%%%%%%%%%
%%%%%%%%%%%%%%%%%%%%%%%%%%%%%%%%%%%%%%%%%%%%%%%%%%%%%%%%%%%%%%%%%%%%%%%%%%%%%%%%%%%%%%%%%%%%%%%%%%%%%%%

\section{Introduction}Originating from the black-hole information paradox, scrambling of quantum information has become a central feature differentiating between integrable systems and chaotic systems \cite{Hayden:2007cs, Sekino:2008he, Shenker:2013pqa}. Several diagnostics have been proposed to identify information scrambling such as the spectral form factor (SSF) \cite{PhysRevE.55.4067, Haake:2010fgh, Cotler:2016fpe}, operator entanglement \cite{Zanardi:2001zza, Alba:2019okd, Bertini:2019gbu, Nie:2018dfe, Styliaris:2020tde}, entanglement spread \cite{Lashkari:2011yi, Asplund:2015eha, Liu:2017lem, PhysRevX.7.031016, PhysRevX.9.021033, Alba:2019ybw}, out-of-time-ordered correlators (OTOC) \cite{OTOC, Shenker:2013pqa, Maldacena:2015waa, Roberts:2014ifa, Hosur:2015ylk} and tripartite mutual information (TMI) \cite{Hosur:2015ylk, Iyoda:2017pxe, Pappalardi:2018frz}.
However, it has recently been shown that observing an information scrambling diagnostic does not necessarily imply quantum chaos in the system. Specifically, non-chaotic/integrable systems have been identified that exhibit the exponential growth of OTOC \cite{PhysRevE.101.010202, Xu:2019lhc, Hashimoto:2020xfr} and operator entanglement \cite{Dowling:2023hqc}.
These recent observations raise the question of how quantum information scrambling, distinct from quantum chaos \footnote{When we refer to chaos, we mean many-body chaos unless explicitly stated as single-particle chaos.}, can be characterized in quantum systems. In particular, identifying the features that drive information scrambling can provide insight into the dynamics of complex quantum systems.

In this paper, we investigate information scrambling in multimode bosonic systems with quadratic Hamiltonians. We consider various diagnostics and examine different amounts of randomness in the Hamiltonian, ranging from the local translational invariant to the completely random non-local models \footnote{The latter model may be seen as bosonic counterparts for the SYK$_2$ model.}. %We investigate the entanglement dynamics in local Hamiltonians. 
For local Hamiltonians and initial Gaussian states, even with a small amount of randomness that preserves locality while removing translational invariance in the Hamiltonian, we observe the disappearance of the memory effect indicating the delocalization of information in the system. We also observe the emergence of negative TMI values for initial squeezed-vacuum states under local and nonlocal Hamiltonians. We find that increasing the amount of randomness leads TMI to rapidly saturate at large negative values indicating strong scrambling in these systems. 

These systems are \textit{quantum integrable} as the number operator associated with each mode commutes with those of other modes and with the Hamiltonian. Therefore, in general, we can see that OTOC and SFF display a power law growth and a non-linear ramp, respectively, distinct from the exponential growth and linear ramp seen in chaotic systems. However, these systems, under certain constraints, \textit{can} exhibit chaotic behavior. Specifically, a decomposition of random Hamiltonians in terms of orthogonal Hermitian operators can have a chaotic component with a Wigner-Dyson energy-level statistics. Hence, we can observe a chaotic behavior when the system is prepared in states within the corresponding subspace of the Hilbert space, spanned by the eigenvectors of the chaotic component. In particular, we can see that the SFF associated with the chaotic component shows a linear ramp, and the OTOC of certain operators that live in this subspace can demonstrate exponential growth.

Our study not only sheds light on new aspects of information scrambling but also offers new insights into the Gaussian dynamics of continuous-variable systems. In particular, these systems are experimentally accessible in quantum optics platforms. Hence, we consider circuit models involving linear optical elements and squeezed-vacuum states, which have been used in Gaussian boson sampling experiments~\cite{doi:10.1126/science.abe8770,Zhong2021,Madsen2022,Deng2023}. Our simulations demonstrate that the information scrambling characteristics of these models can be experimentally probed. Furthermore, our results can be used to investigate the role of randomness in the dynamics of such systems. Remarkably, these characteristics persist even when noise added to the initial states in these models destroys entanglement.

This paper is structured as follows. In Section~\ref{sec:setup}, We explain the setup and introduce random quadratic Hamiltonian models that we use. In Section~\ref{Sec:memory}, we investigate the memory effect in Gaussian dynamics induced by local quadratic Hamiltonians. We consider TMI in Section~\ref{sec:tmi}. In Section~\ref{sec:chatoic-comp}, we show that  random quadratic Hamitonians can have a chaotic component. In Section~\ref{sec:renyi-2}, we show that using the R\'enyi-2 entropy instead of the von Neuman entropy leads to similar results for information scrambling in terms of the memory effect and TMI. The paper is concluded in Section~\ref{sec:disc} and supplemented by three appendices.

%%%%%%%%%%%%%%%%%%%%%%%%%%%%%%%%%%%%%%%%%%%%%%%%%%%%%%%%%%%%%%%%%%%%%%%%%%%%%%%%%%%%%%%%%%%%%%%%%%%%%%%
\section{Setup}\label{sec:setup} We begin by briefly reviewing the system and describing the random quadratic Hamiltonians used in our analysis; further details are provided in Appendix A. We consider $N$-mode bosonic systems described by the vector of quadrature operators $r=(q_1,\dots,q_N,p_1,\dots,p_N)^\top$, satisfying the canonical commutation relations. Quantum states that can be fully described by the first-order moments $\langle r\rangle=\tr(\rho r)$ and the covariance matrix with the matrix elements $\bm\sigma_{jk}=\la r_j r_k+r_k r_j\rangle-2\la r_j\lb \la r_k\lb$ are known as Gaussian states~\cite{WeedbrookRMP2012,Serafini17}. If a covariance matrix satisfies $\bm\sigma-\mathbb{1}_{2N}\geq0$, where $\mathbb{1}_{2N}$ is the $2N\times 2N$ identity matrix, the corresponding state is called {\it classical} in quantum optics, and can be expressed as a statistical mixture of product coherent states.

We consider quadratic Hamiltonian of this form
\begin{equation}\label{eq:H-M}
H=\frac{1}{2}r^\top \bm M r
\;,\;\;\;\;\bm M=\begin{pmatrix}
    \bm M_{q} & \bm M_{qp} \\  \bm M_{qp}^\top & \bm M_{p}
\end{pmatrix}
\end{equation}
with matrix $\bm M$ being real, symmetric, and positive definite. These Hamiltonians generate Gaussian unitaries that preserve Gaussianity of states. If $\bm M$ can be diagonalized using a symplectic, orthogonal matrix such that $\bm M_{q}=\bm M_{p}$, the corresponding unitary is known as {\it passive} and preserve the number of particles \footnote{The set of these passive transformations is isomorphic to the compact group U($N$), making uniform sampling with a Haar measure well-defined.}. Passive unitaries linearly transform the creation operators $U_\text p^\dagger a^\dagger_j U_\text p=\sum_n  \bm U_{jn} a^\dagger_n $, where $\bm U_{jn}$ are the elements of an $N\times N$ unitary transfer matrix $\bm U$. In quantum optics, these transformations can be realized using linear-optical networks. A special 2-mode linear-optical network is a beam splitter (BS) with a 2$\times$2 unitary matrix $\bm U_{\rm BS}$. Gaussian unitaries that do not preserve particle numbers are called {\it active}. An example is single-mode squeezing unitary defined by  $U_{\text{sq},\lambda}^\dagger (q,p) U_{\text{sq},\lambda}=(q e^{-\lambda},p e^{\lambda})$, transforming the vacuum state into the squeezed-vacuum state with the covariance matrix $\text{diag}(e^{-2\lambda}, e^{2\lambda})$, where $\lambda$ is the squeezing parameter.

Let us first consider a one-dimensional chain with the nearest-neighbor couplings and periodic boundary conditions. We chose $\bm M_{qp}=0$, $\bm M_{p}=\mathbb{1}_N/\epsilon$, and $\bm M_{q}=1/\epsilon\; \mathrm{circ}(\epsilon^2 m^2+2,-1,\cdots,-1)$. This Hamiltonian known as the Harmonic Lattice model (HL) at large-$N$ describes a Klein-Gordon free massive scalar field in two-dimensional spacetime, regularized on a lattice with the spacing $\epsilon$ and mass $m$. To recover the continuum limit, we have to take the limit of $\epsilon\to0$. We, however, consider a \textit{non-vanishing} lattice spacing, corresponding to a UV cut-off in the field theory description. For convenience, we set $\epsilon=1$, and this Hamiltonain is denoted by $H_{\mathrm{HL}}$.

%%%%%%%%%%%%%%%%%%%%%%%%%%%%%%%%%%%%%%%%%%%%%%%%%%%%%%%%%%%%%%%%%%%%%%%%%%%%%%%%%%%%%%%%%%%%%%%%%%%%%%%

We now describe our models of random Gaussian dynamics to investigate information scrambling. The first family of models denoted by type-I are active random Hamiltonians. Model (Ia) is a \textit{disordered} version of the local HL model \be\label{eq:HLrandom}
H_{\mathrm{DHL}}=\frac{1}{2}\sum_{n=1}^N\left[p_n^2+ m^2q_n^2+J_n\left(q_{n+1}-q_n\right)^2\right].
\ee
where $J_n$ are random real numbers. We denote this \textit{local} Hamiltonian with random interactions by the Disordered Harmonic Lattice (DHL) Hamiltonian.
In addition to this random local model, we consider non-local models (Ib) where $\bm M=\text{diag}(\bm M_q,\mathbb{1}_N)$, where $\bm M_q$ is chosen from either the Gaussian orthogonal ensemble (GOE) or the Gaussian unitary ensemble (GUE).

% \begin{figure}[t]
% \begin{center}
% \includegraphics[scale=0.5]{circ13.pdf}
% \hspace{8mm}
% \includegraphics[scale=0.5]{circ23.pdf}
% \end{center}
% \caption{Random dynamics generated by passive linear-optical circuits. In model (IIa), a two-layer circuit consisting of random beam splitters is applied at each time step. In model (IIb)  }
% \label{fig:LON-circuits}
% \end{figure}

The second family of models is defined in terms of linear-optical networks generating random Gaussian dynamics. We consider a passive local circuit model (IIa) involving two layers of BSs described by $\otimes_{j=1} U_{\text{BS},2j} \otimes_{j=1} U_{\text{BS},2j-1}$ act on the $N$-mode state in each time step. We assume periodic boundary conditions, and the BS operator $U_{\text{BS},j}$ acts on $j$th and $(j+1)$th modes. We choose the BS transfer matrix $\bm U_{\text{BS}}$ from a Haar measure to generate random dynamics. Note that in this model, as we show in the following section, the use of fixed balanced 50:50 BSs corresponds to a circuit model representing HL-like evolution. 

%We also consider a passive non-local circuit model (IIb), involving an $N$-mode linear-optical network whose $N\times N$  unitary transfer matrix $\bm U$ is chosen from a Haar measure. 

%\Saleh{We also consider a passive non-local circuit model (IIb) by randomly selecting the matrix $M$ in Eq.~\eqref{eq:H-M} from the set of positive, symmetric matrices that can be diagonalized using symplectic and orthogonal matrices. Specifically, we do this by considering Hamiltonians that can be expressed as $H=U_\text p  H_\text D U^{\dagger}_\text p$, where $H_\text D=\!\sum_{j=1}^N \omega_j \big(q_j^2 +p_j^2\big)\!$ with $\{\omega_j\}$ being randomly distributed, and the passive unitary $U_\text p$ is chosen such that the corresponding $N\times N$  transfer matrix $U$ is Haar-random. In this model, the random passive unitary time evolution is given by $V(t)=U_\text p \,e^{-i H_\text D t}\, U^{\dagger}_\text p$, which can be described by transfer matrix $\bm V(t)=\bm U\cdot \mathrm{diag}(e^{-i \omega_k t})\cdot\bm U^{\dagger}$.  For more details about passive Hamiltonians and matrix $\bm V(t)$ see Appendix C. }

We also consider a passive non-local circuit model (IIb). This model is defined by a random Hamiltonian of the form $H=U_\text p  H_\text D U^{\dagger}_\text p$, where $H_\text D=\!\sum_{j=1}^N \omega_j \big(q_j^2 +p_j^2\big)\!$ contains randomly distributed frequencies $\{\omega_j\}$. The passive unitary $U_\text p$ is generated such that the corresponding $N\times N$  transfer matrix, $\bm U$, is Haar-random. Note that this construction means that the matrix $M$ in Eq.~\eqref{eq:H-M} is positive, symmetric, and diagonalizable by symplectic and orthogonal matrices. The resulting time evolution is given by the unitary $V(t)=U_\text p \,e^{-i H_\text D t}\, U^{\dagger}_\text p$, which has the transfer matrix $\bm V(t)=\bm U\cdot \mathrm{diag}(e^{-i \omega_k t})\cdot\bm U^{\dagger}$. Further details on passive Hamiltonians and this transfer matrix are provided in Appendix C.

\begin{figure}[t]
\begin{center}
\includegraphics[scale=0.4]{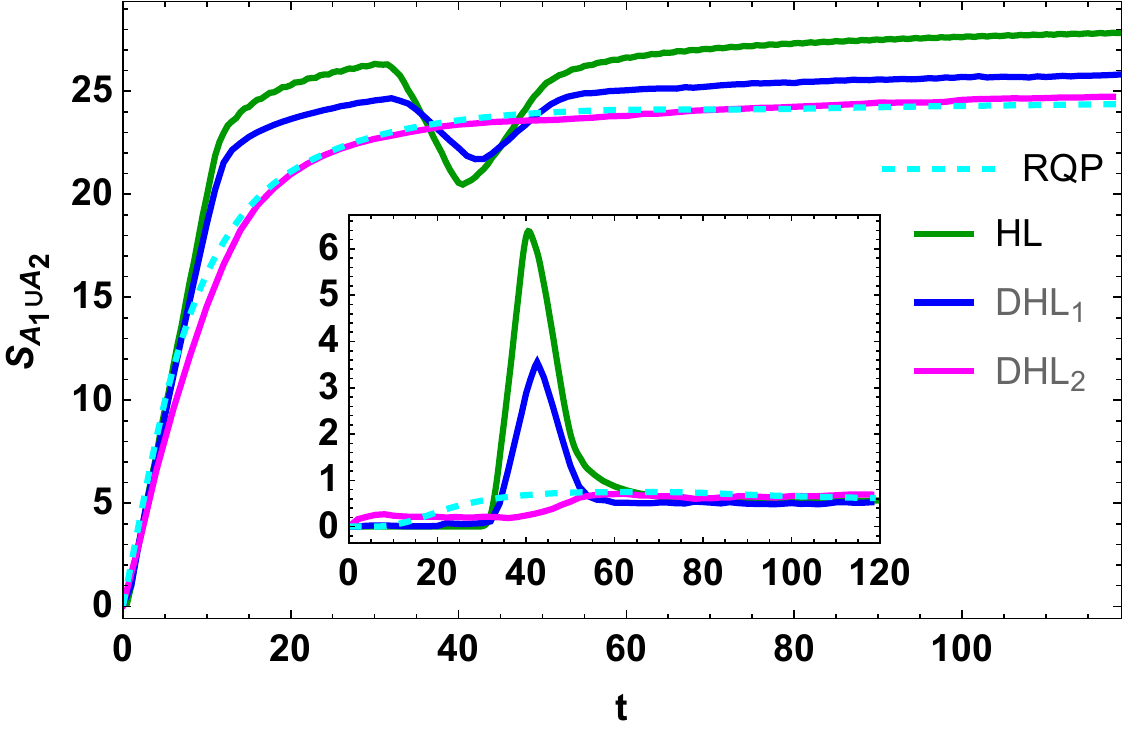}
\end{center}
\caption{The disappearance of the memory effect, the dip in the time evolution of the joint entropy $S_{A_1\cup A_2}$ in the presence of randomness in the dynamics. $S_{A_1\cup A_2}(t)$ in HL versus DHL (Ia) model. The ground state of $H_{\mathrm{HL}}$ with $m=2$, denoted by $\ket{\psi_{\text{HL},0}}$ is evolved by unitary operators associated with $H_{\mathrm{HL}}$ and $H_{\mathrm{DHL}}$ with $m=10^{-7}$ (such a process is known as a mass quench). We set the number of modes in the disjoint blocks $N_{A_1}=N_{A_2}=20$, the separation between them $N_d=60$, and the system size $N=500$. DHL$_1$: $J_n\in(0.5,1.5)$ and DHL$_2$: $J_n\in(0,2)$ from a uniform distribution. The inset shows the spike in MI in HL dynamics, while the spike effectively disappears due to the delocalization of information in DHL dynamics. The dashed cyan curves show random-quasi-particle prediction for $s(\kappa)=0.32$ and $v(\kappa)$ chosen from exponential distribution $e^{-1.5\kappa}$ shifted by a constant 0.15 and averaged over 50 samples.}
\label{fig:SAB}
\end{figure}

\begin{figure}[t]
\begin{center}
\includegraphics[scale=0.6]{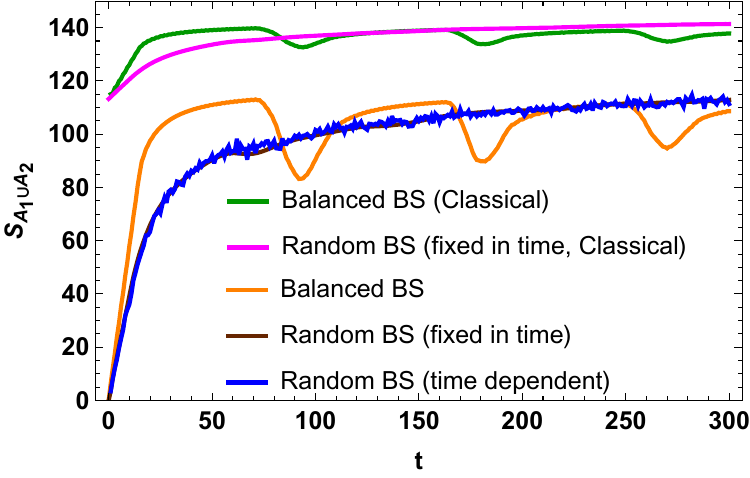}
\end{center}
\caption{$S_{A_1\cup A_2}$ in the passive circuit model (IIb). The initial state is the tensor product of single-mode squeezed vacuum states with $\lambda_i=2$. We set $N_{A_1}=N_{A_2}=40$ and $N_d=200$. We consider two scenarios: the BS transfer matrix $\bm U_{\mathrm{BS}}$ is chosen randomly at every single time step, and $\bm U_{\mathrm{BS}}$ is fixed during the evolution. For a balanced BS, we observe memory effects, where the second and third dips in the orange curve correspond to entanglement revivals. We observe a similar effect if we add vacuum noise to the squeezed-vacuum states to make the initial state classical and destroy entanglement in the states. All random results are averaged over 300 samples.}
\label{fig:SABcircuit}
\end{figure}

%%%%%%%%%%%%%%%%%%%%%%%%%%%%%%%%%%%%%%%%%%%%%%%%%%%%%%%%%%%%%%%%%%%%%%%%%%%%%%%%%%%%%%%%%%%%%%%%%%%%%%%

\section{Memory Effect During Entanglement Dynamics}
\label{Sec:memory}
The pattern of entanglement dynamics of disjoint blocks is a distinct measure of information scrambling. Here we denote these blocks by $A_1$ and $A_2$ separated by $d$ with $d>\ell_{A_1},\ell_{A_2}$ (with unit lattice spacing, length $\ell_{A_i}$ is equal to the number of modes $N_{A_i}$, and $d=N_{d}$). Characterizing this measure originates in a holographic viewpoint wherein chaotic systems are realized by holographic conformal field theories (CFTs) \footnote{A holographic CFT is defined by a large number of degrees of freedom (large central charge in 2$d$) and a sparse spectrum, i.e. a large gap between low spin and high spin operators.}. The joint entanglement entropy $S_{A_1\cup A_2}(t)$ in integrable systems (e.g. free CFTs) shows a long-term \textit{memory effect}, namely a dip-ramp occurring after saturation, that is absent in chaotic systems (e.g. holographic CFTs) \cite{Asplund:2015eha}. Such a behavior has been realized in a wide spectrum of systems \cite{Allais:2011ys, Balasubramanian:2011at, Asplund:2013zba, Leichenauer:2015xra, Asplund:2015eha, PhysRevX.7.031016, PhysRevX.9.021033, PhysRevB.101.094304, Alba:2019ybw}. This effect indicates that initially uncorrelated $A_1$ and $A_2$ become correlated during the period of dip-ramp formation \cite{Coser:2014gsa}.

Here we consider the ground state of $H_{\mathrm{HL}}$ with $m\neq0$, denoted by $\ket{\psi_{\text{HL},0}}$, as the initial state. Figure \ref{fig:SAB} presents our simulation results for the joint entropy during the unitary evolution of $\ket{\psi_{\text{HL},0}}$ induced by $H_{\mathrm{HL}}$ with $m \to 0$ (corresponding to a free bosonic CFT with unit central charge), which shows an apparent memory effect. In contrast, the evolution under $H_{\mathrm{DHL}}$ with the same $m$ shows that, for a small amount of randomness, $J_n \in (0.5, 1.5)$, the dip corresponding to the memory effect becomes shallower, and for $J_n \in (0, 2)$, there is no observable memory effect. In the inset the corresponding mutual information (MI), $I_2(A_1:A_2)=S_{A_1}+S_{A_2}-S_{A_1\cup A_2}$, is shown to compare the spike corresponding to the $H_{\mathrm{HL}}$ dynamics and the delocalization of information due to random local $H_{\mathrm{DHL}}$ dynamics. 

The evolution of entanglement in nonrandom integrable systems can be understood in terms of free streaming quasi-particles \cite{Calabrese:2005in, Alba:2017ekd}. In this picture, entanglement spreads via pairs created throughout the system. Each pair is initially localized and becomes shared between regions as the particles propagate ballistically. The entanglement entropy of a region grows when one member of a pair enters while its partner remains outside.

The integrable nature of our models suggests the existence of a quasi-particle description. However, the inherent randomness prevents the definition of a dispersion relation for the Hamiltonian, and thus, it is not possible to read the precise contribution of each decoupled mode and employ a predictive quasi-particle formula as in \cite{Alba:2017ekd, Alba:2017lvc}. We rather propose a heuristic quasi-particle formula for this model: for a region with length $\ell_A$, the entanglement entropy similar to \cite{Alba:2017ekd} follows  
\be
S_A=2t\sum_{2t|v(\kappa)|<\ell_A}v(\kappa)s(\kappa)+\ell_A\sum_{2t|v(\kappa)|>\ell_A}s(\kappa)\,.
\ee
Here $\kappa$ solely labels the normal modes (not referring to quasi-momenta as it does in translational invariant nonrandom systems), $v(\kappa)$ and $s(\kappa)$ are chosen from a random distribution. Using a straightforward generalization of this formula for disjoint blocks similar to \cite{Alba:2018hie}, we find a qualitatively nice agreement with our numerical results depicted in Fig.~\ref{fig:SAB} denoted by the dashed cyan curve. This qualitative behavior is largely independent of the specific choices of distributions for $v(\kappa)$ and $s(\kappa)$.

We find similar behavior with the passive circuit model (IIa) with single-mode squeezed vacuum input states; see Fig.~\ref{fig:SABcircuit}. Note that in an integrable periodic system, the correlation between $A_1$ and $A_2$ revives periodically, known as entanglement revivals \cite{Bertini:2018fbz, Piroli:2019umh, Chan:2017kzq}. This effect can be observed in Fig.~\ref{fig:SABcircuit} for a balanced BS with $\bm U_{\text{BS}}=(\mathbb{1}_2+i\bm X)/\sqrt{2}$, where $\bm X$ is the $X$-Pauli matrix, while the memory effect disappears for the case of random BSs as another sign of scrambling in these systems.

Interestingly, we observe a diminished version of similar features in Fig.~\ref{fig:SABcircuit}, even with the addition of one unit of vacuum noise to the initial squeezed-vacuum states in model (IIa). In this case, the initial state is the tensor product of states with the covariance matrix $\text{diag}(e^{-2\lambda}+1, e^{2\lambda}+1)$, which are classical and can be viewed as statistical mixtures of coherent states. As coherent states traverse through linear-optical networks like classical waves, no entanglement is generated at the output~\cite{InSitu2020}.  

\begin{figure*}[t!]
\begin{center}
\includegraphics[scale=0.37]{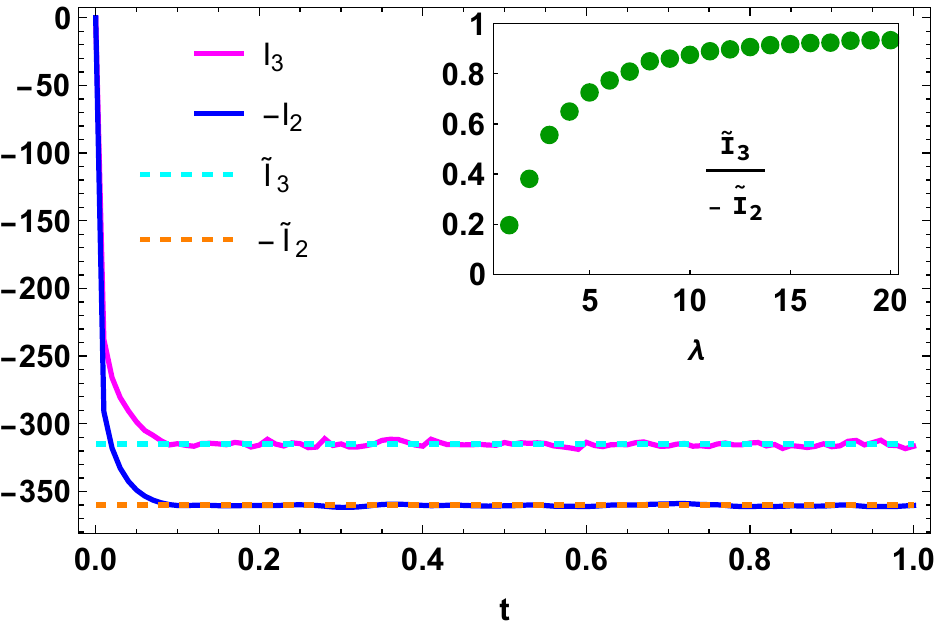}
%\hspace{1mm}
\includegraphics[scale=0.37]{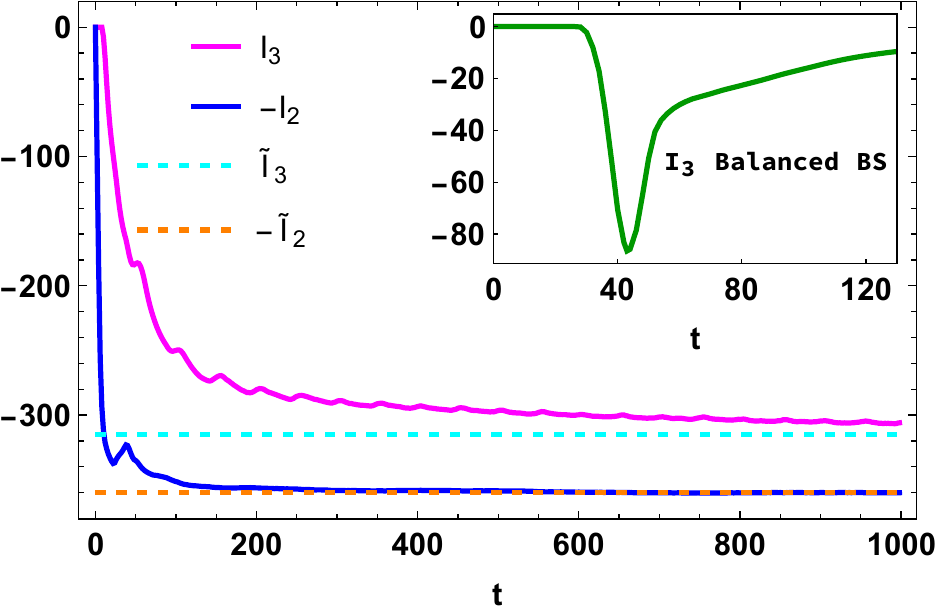}
%\hspace{1mm}
\includegraphics[scale=0.37]{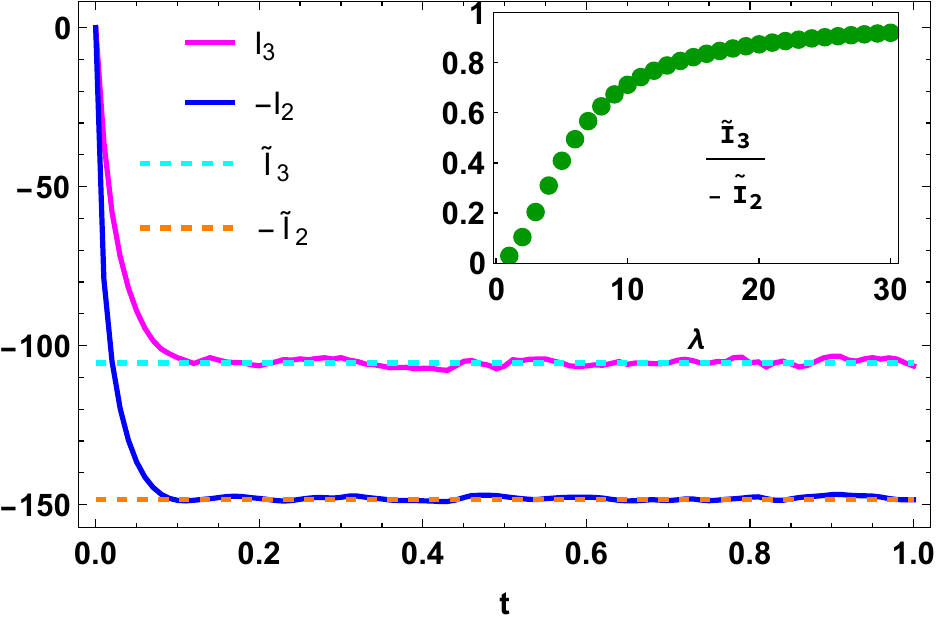}
\end{center}
\caption{
TMI time evolution by passive random unitaries. The initial state is the tensor product of squeezed-vacuum states with $\lambda=10$, and we set $N=200$ with $N_A=20$, $N_B=80$, and $N_C=30$. \textit{Left:} A random non-local passive evolution $V(t)$ (a single sample). TMI instantly approaches and saturates to $\tilde{I}_3$ (dashed cyan line), the value obtained for TMI for the same state evolved (a single time step) by a Haar random unitary (IIb). The inset shows the ratio of the saturation value of $\tilde{I}_3/(-\tilde{I}_2)$ approaches to unity as $\lambda$ increases.
\textit{Middle:} random local evolution for  $N_{d_{AB}}=0$ and $N_{d_{BC}}=70$, averaged over 500 samples. In this case, $I_3$ asymptotically approaches the same value of $\tilde{I}_3$ in the left panel.
Note that the saturation timescale for the local evolution is significantly larger than the nonlocal evolution. The inset shows the evolution of TMI with the (non-random) balanced BS where TMI does not saturate to a negative value (here, the system size is enlarged to $N=500$ to postpone entanglement revival effects). \textit{Right:} TMI time evolution by a passive random unitary with the initial state being the tensor product of classical states with the covariance matrix $\text{diag}(e^{-20}+1, e^{20}+1)$. Here there is no entanglement in the system. The configuration is the same as the right panel.
}
\label{fig:I3}
\end{figure*}

%%%%%%%%%%%%%%%%%%%%%%%%%%%%%%%%%%%%%%%%%%%%%%%%%%%%%%%%%%%%%%%%%%%%%%%%%%%%%%%%%%%%%%%%%%%%%%%%%%%%%%%
\section{Tripartite Mutual Information}
\label{sec:tmi}
Tripartite mutual information is defined in terms of MI as 
\be\label{eq:I3}
I_3(A:B:C)\!=I_2(A:B)+I_2(A:C)-I_2(A:B\cup C).
\ee
The sign of $I_3$ is not definite in general \cite{Casini:2008wt, Rangamani:2015qwa, Rota:2015wge, Agon:2021lus} though in holographic theories we always have $I_3\leq 0$ \cite{Hayden:2011ag}. While MI is non-negative due to the subadditivity property of von-Neumann entropy, the most negative value for  $I_3$ arises from its final term. When this property is applied to a careful partitioning of the system, it is capable of measuring scrambling, as introduced in \cite{Hosur:2015ylk}. Suppose we divide our $N$-mode system into four parties denoted by $A,B,C$, and $D$ where $N_A,N_C<N/4$, while $N/4<N_B<N/2$ such that $N_B+N_C>N/2$. When the system undergoes a strong scrambling unitary evolution, since non of the parties are larger than half of the system, we expect $I_2(A:B)$ and $I_2(A:C)$ to be very small compared to $I_2(A:B\cup C)$ due to $N_B+N_C>N/2$. In other words, for such a configuration, the stronger the scrambling, the closer the value of $I_3$ approaches its most negative possible value. This property, discussed in the framework of quantum channels in \cite{Hosur:2015ylk}, is directly applicable to evaporating black holes.

%\textcolor{blue}{We consider the time evolution of $I_3$ in a uniformly squeezed state evolved by passive random unitaries in models (IIa) and (IIb), where a well-defined \textit{Haar} measure applies. For the case of model (IIa), similar to section \ref{Sec:memory}, the time evolution is generated by the successive application of the two-layer circuit. For the case of model (IIb), since the unitary operates on the full system in a single time step, we introduce a heuristic passive random unitary time evolution operator as $V(t)=U_\text p \,e^{-i H_\text D t}\, U^{\dagger}_\text p$, where $H_\text D=\!\sum_{j=1}^N \omega_j \big(a^\dagger_j a_j+\frac{1}{2}\big)\!$.} The corresponding transfer matrix of this operator takes the form of $\bm V(t)=\bm U\cdot \mathrm{diag}(e^{-i \omega_k t})\cdot\bm U^{\dagger}$ where $\bm U$ is a Haar random unitary transfer matrix associated with $U_\text p$, and $\{\omega_k\}$ are chosen from an arbitrary random distribution; see Appendix C for more details about $\bm V(t)$ and Appendix B for similar discussion for active models (Ia) and (Ib).

We consider the time evolution of $I_3$ in a uniformly squeezed state evolved by passive random unitaries in models (IIa) and (IIb), where a well-defined \textit{Haar} measure applies. The time evolution of $I_3$ under models (Ia) and (Ib) is discussed in Appendix B.

In Fig.~\ref{fig:I3}, we present the simulation results for these passive random evolutions of TMI for an initial uniform squeezed-vacuum state, $\ket{\psi_0}=\otimes_{j=1}^{N} U_{\text{sq},\lambda}\ket{\bm 0}$ with $\ket{\bm 0}=\ket{0,\dots,0}$ being the vacuum state of $N$ modes. The left panel shows the nonlocal case where the dynamics is driven by $V(t)$ in model (IIb). In this case, TMI quickly takes a negative value and saturates to the value corresponding to $U_\text p\ket{\psi_0}$ denoted by $\tilde{I}_3$. We similarly denote the value of $I_2(A:B\cup C)$ corresponding to $U_\text p\ket{\psi_0}$ by $\tilde{I}_2$. The inset of the right panel shows that as the entanglement in the initial state (the squeezing parameter $\lambda$) increases, $\tilde{I}_3\to -\tilde{I}_2$ indicating that \textit{highly entangled states undergo strong scrambling under passive random Gaussian unitary evolution}.
This behavior aligns with the findings of \cite{Iosue:2022lnb}, which demonstrated that increasing $\lambda$ brings the Page curve of these random Gaussian states closer to that of typical random states.

In the middle panel of Fig.~\ref{fig:I3}, we present the dynamics driven by the \textit{local} circuit model (IIa). The random local dynamics cause the TMI to take negative values, typically considered an indication of scrambling \cite{Iyoda:2017pxe, Pappalardi:2018frz, Seshadri:2018yya, Schnaack:2018bhs, PhysRevB.108.134434, LoMonaco:2023xws}. It is important to note that simply taking on a negative value does not indicate scrambling; the negative value must also be non-increasing. As shown in the inset, under \textit{non-random} local dynamics, TMI takes negative values but does not saturate, instead starting to increase after some time. Here in the main panel, in contrast to nonrandom local dynamics, the behavior for random local dynamics is very similar to the nonlocal case, in that TMI ultimately converges to the same value of $\tilde{I}_3$, albeit over a much longer timescale. After adding a unit shot noise to $\ket{\psi_0}$, we find the same qualitative behavior for TMI as illustrated in the right panel of Fig.\ref{fig:I3}. 

%%%%%%%%%%%%%%%%%%%%%%%%%%%%%%%%%%%%%%%%%%%%%%%%%%%%%%%%%%%%%%%%%%%%%%%%%%%%%%%%%%%%%%%%%%%%%%%%%%%%%%%
\section{Chaotic component of random quadratic Hamiltonains}\label{sec:chatoic-comp}
In general, using a Gaussian unitary $U_G$, any quadratic Hamiltonains can be transformed to an uncoupled Hamiltonian and can be decomposed as~\cite{Serafini17}
\begin{equation}\label{eq:Hdecomposition}
    H=U_G \bigg(\!\sum_{j=1}^N \omega_j \big(a^\dagger_j a_j+\frac{1}{2}\big)\!\bigg) U_G^\dagger=  \sum_{n=0}^\infty U_G \Pi_n U_G^\dagger, 
\end{equation}
where $\{\omega_j\}$ are the symplectic eigenvalues of $\bm M$ in Eq.(\ref{eq:H-M}) and
$\{\Pi_n\}$ are orthogonal Hermitian operators acting on the $n$-particle subspace of the Hilbert space; for example, $\Pi_0=\frac 12 (\omega_1+\dots+\omega_N)\ket{\bm 0}\bra{\bm 0}$, and $\Pi_1=\frac 32 \sum_{j=1}^N\omega_j a^\dagger_j\ket{\bm 0}\bra{\bm 0}a_j$. Note that the Hilbert space of the $N$-mode system can be decomposed as $\mathcal{H}=\bigoplus_n\tilde{\mathcal{H}}_n$, where the subspace $\tilde{\mathcal{H}}_n$ can be spanned by the eigenvectors of $\tilde\Pi_n=U_G \Pi_n U_G^\dagger$. We can see that if the state of the system is $\ket{\phi_n}\in \tilde{\mathcal{H}}_n$, the dynamics is governed by the Hamiltonian component $\tilde\Pi_n$. Therefore, if 
the energy spectrum of $\tilde\Pi_1$, $\{\omega_j\}$, obey a Wigner-Dyson statistics, for states in the $N$-dimensional subspace $\tilde{\mathcal{H}}_1$, spanned by the basis $\{U_G a^\dagger_j\ket{\bm 0}\}_{j=1}^N$, the system can display chaotic behavior. Note that states $\ket{\phi_1}\in\tilde{\mathcal{H}}_1$ are non-Gaussian, and if $U_G$ is passive, $\tilde{\mathcal{H}}_1$ is the single-particle subspace. Note also that states and unitary evolution of any quantum system with $N$-dimensional Hilbert space can be realized using this system.

In the following subsections, we show that OTOC for local operators grows polynomially in time, and SFF exhibits a non-linear ramp. However, for states and local operators in  $\tilde{\mathcal{H}}_1$ OTOC grows exponentially and SFF  corresponding to this sector exhibits a linear ramp, as expected in chaotic systems.

\subsection{Out of Time-Ordered Correlators}
OTOC has been defined by quantizing the classical observation that the sensitivity to the initial conditions can be quantified by the Poisson bracket $\{q(t),p\}$ \cite{OTOC, Shenker:2013pqa, PRXQuantum.5.010201}. The exponential growth of OTOC was known to indicate quantum chaos~\cite{Shenker:2013pqa}; however, an unstable saddle point in integrable systems can also lead to the exponential growth of OTOC \cite{Xu:2019lhc}. A natural question is how OTOC behaves in integrable systems in the presence of randomness.

\begin{figure}[h]
%\begin{center}
\includegraphics[scale=0.4]{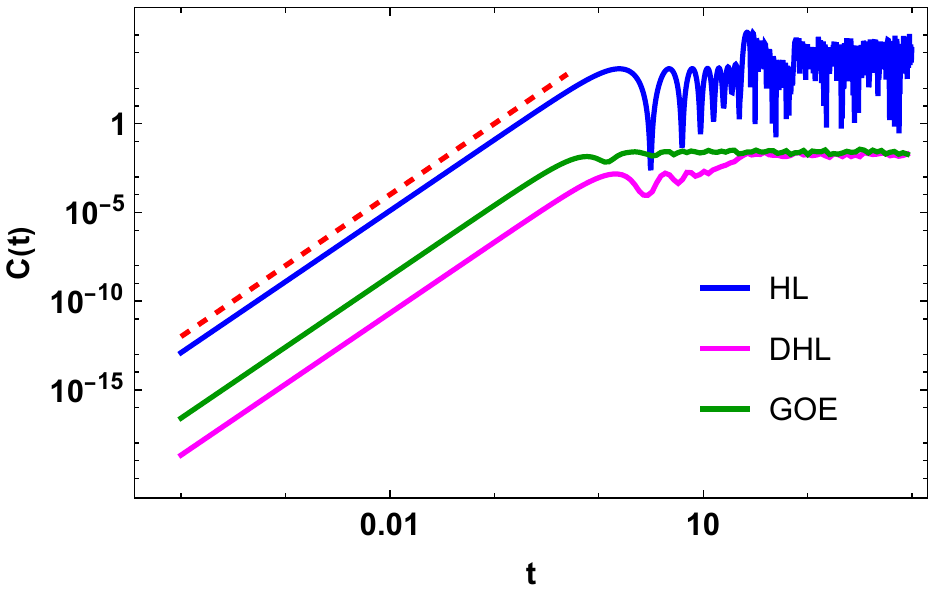}
%\end{center}
\caption{OTOC for $q_j(t)$ and $p_k$ in the ground state of the HL, DHL (Ia) and $\bm M_q\in$GOE (Ib), showing a power law growth. We set $j=1$, $k=N/2$, and $N=100$. The DHL and GOE results are averaged over 100 samples. The dashed red line $\propto t^4$.}
\label{fig:otoc}
\end{figure}

Here we consider OTOC in terms of the canonical operators with respect to a thermal state
\begin{equation}
C_{\beta,jk}(t)=-\tr\!\left(\frac{e^{-\beta H}}{\tr\left(e^{-\beta H}\right)} \big[q_j(t),p_k\big]^2\right)
\end{equation}
where $q_j(t)=e^{itH}q_je^{-itH}$ and $j,k$ stands for the mode index. In Appendix C, we show that if the Hamiltonian matrix can be diagonalized by a symplectic transformation of the form $\bm S=\bm S_q\oplus \bm S_p$, OTOC on the the ground state is given by 
\begin{align}
\begin{split}\label{eq:C1}
C_{\infty,jk}(t)&=\big(\bm S_q\, \mathrm{diag}(\cos(\omega_1 t),\cdots,\cos(\omega_N t)) \bm S_p^\top\big)_{jk}^2,
\end{split}
\end{align}
where $\beta\to\infty$ implies zero temperature. Using $\bm S_q \bm S_p^\top=\mathbb{1}_N$, one can easily verify that $C_{\infty,jk}(t\ll 1)\sim t^4$ for $k\neq j$ and $C_{\infty,jk}(t\ll 1)\sim t^2$ for $k=j$.  Our numerical simulations for models (Ia) and (Ib) presented in Fig. \ref{fig:otoc} agree with this results \footnote{See \cite{Garcia-Garcia:2024mdv} where power-law behavior for OTOC in SYK$_2$ is reported.}. 

We analyze a generic form of OTOC for passive Gaussian unitaries in Appendix C. At zero temperature, this quantity depends on a single time-dependent parameter, whose saturation time decreases with the number of modes, as $t_*\sim1/\log N$. This feature results in shortening the power law growth in large systems followed by a plateau. For OTOC under non-Gaussian dynamics of continuous variable systems, see \cite{Zhuang:2019jyq}.

As discussed, the unitary evolution of states of an $N$-mode bosonic system in the $N$-dimensional subspace $\tilde{\mathcal{H}}_1$ is described by $\tilde\Pi_1=\frac 32 \sum_{j=1}^N\omega_j \ket{j}\bra{j}$ component of the quadratic Hamiltonian~\eqref{eq:Hdecomposition}, where $\ket{j}=U_G a^\dagger_j\ket{\bm 0}$. Hence, by appropriately choosing $\{\omega_j\}$, this system can simulate any $N$-dimensional quantum system. For instance, with $N=2^n$, one can simulate systems with $n$ spin-1/2 particles. In particular, we know chaotic spin systems under mixed-field Ising dynamics whose Hamiltonians exhibit Wigner-Dyson level statistics~\cite{PhysRevLett.106.050405} and their OTOC grows exponentially~\cite{Roberts:2014isa, PRXQuantum.5.010201}. Therefore, by considering the operators used in the OTOC expressions for these systems and choosing $\{\omega_j\}$ accordingly, we can find OTOCs in terms of states and operators in $\tilde{\mathcal{H}}_1$ that grow exponentially.     

\begin{figure*}
\begin{center}
\includegraphics[scale=0.5]{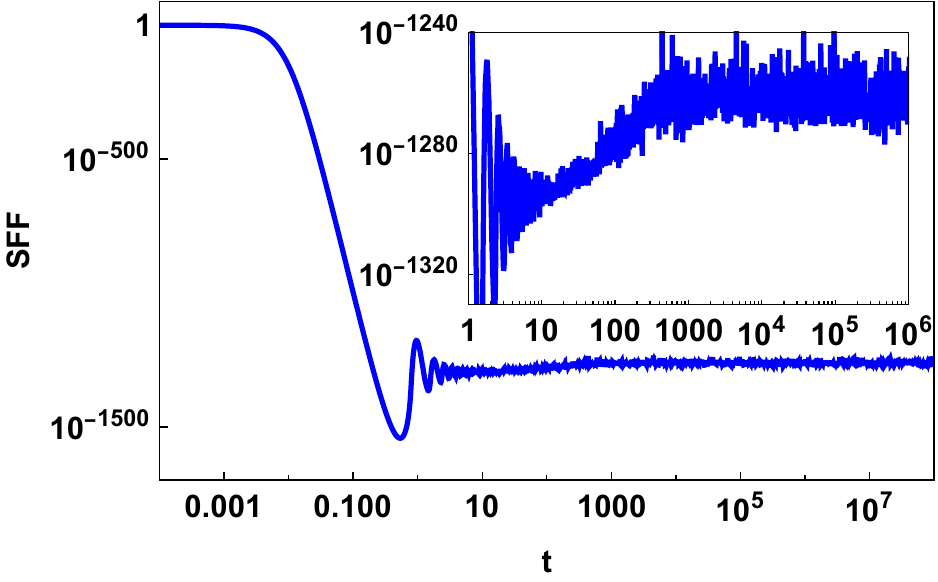}
\hspace{7mm}
\includegraphics[scale=0.5]{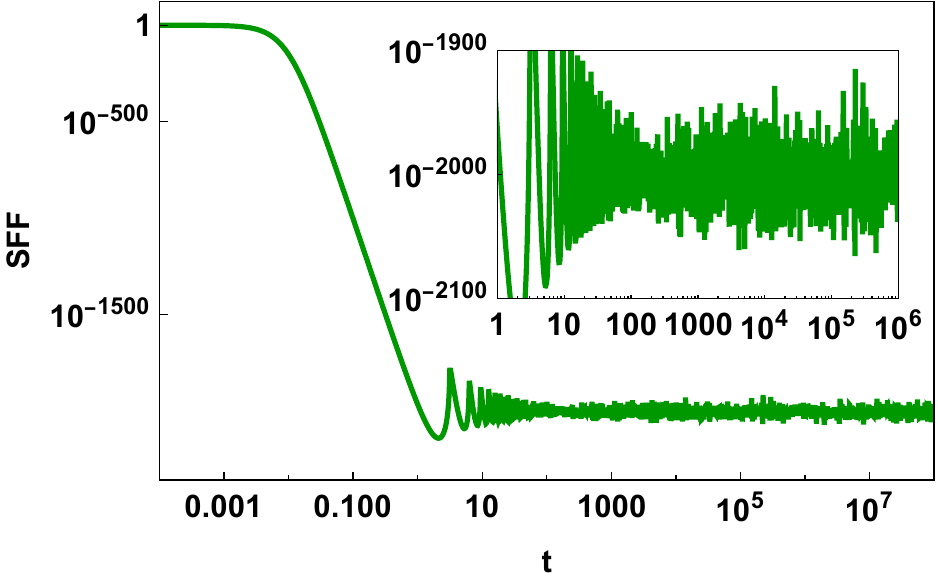}
\end{center}
\caption{The spectral form factor for (left) model (Ib) with $\bm M_{q}\in$ GOE and (right) the HL model (no randomness). The insets is a zoomed version of the same plot. The existence of the ramp in the (Ib) model is clear from comparison of the insets. We set $\beta^{-1}=100$ and the system size $N=500$ for both plots and the (Ib) model result is averaged over 300 samples.}
%\caption{The spectral form factor (SFF) for model (Ib) is shown with $\bm M_{q}\in$ GOE, presenting results for both the entire spectrum (main panel) and the chaotic sector (inset). In both cases, we observe an initial dip followed by a ramp. The inset distinctly exhibits a linear ramp (with the magenta curve scaling as $\propto t^{0.75}$), while the main panel displays a non-linear ramp. To clearly distinguish the ramp from oscillatory behavior, we set $\beta^{-1}=2$ in the main panel and $\beta^{-1}=10$ in the inset. The system size is $N=500$, averaged over 200 samples.}
\label{fig:SFF}
\end{figure*}

\subsection{Spectral Form Factor}
The SFF tracks correlations between energy levels across the entire spectrum of the system over time and exhibits universal behavior in chaotic systems, characterized by a dip-ramp-plateau structure. While chaotic systems display a linear ramp at intermediate times, integrable systems do not show a ramp. The next natural question about our models is how SFF behaves in these systems. Note that the SYK$_2$ model is known to exhibit an exponential ramp \cite{Liao:2020lac, Winer:2020mdc}.

In bosonic quadratic Hamiltonians, the SFF is defined for the $k$th normal mode (the $k$th mode in the diagonal Hamiltonian $H_{\text D}$) in terms of the non-normalized single-mode partition function $Z_k(\beta)=\sum_{n_k} e^{-\beta\omega_k n_k}$ as 
\begin{align*}
g_k(\beta,t)
&=\frac{|Z_k(\beta +i t)|^2}{|Z_k(\beta)|^2}
%\\&
=\frac{\cosh (\beta \omega_k)-1}{\cosh (\beta\omega_k)-\cos (\omega_k t)},    
\end{align*}
and SFF for the entire system is given by $g(\beta,t)=\Pi_k g_k(\beta,t)$. We numerically average $g(\beta,t)$ over thermal states, sometimes called as the quenched quantity.

In this part we focus on model (Ib), where the spectrum of its chaotic sector%, i.e. its single-particle sector, 
obeys a Wigner-Dyson statistics. If we attribute an SFF to the chaotic sector corresponding to the component $\tilde\Pi_1$ defined in Eq.~\eqref{eq:Hdecomposition}, the corresponding SFF exhibits a linear ramp.

In Fig. \ref{fig:SFF}, we compare the behavior of the SFF in model (Ib), with $\bm{M}_q \in$ GOE, to that of the harmonic lattice model with no randomness. A key distinction between the two is the presence of a chaotic sector in model (Ib), which is absent in the harmonic lattice model. The insets in both panels focus on the regime near the onset of the plateau. Our numerical results reveal a clear ramp in model (Ib), emerging at intermediate times and extending up to $t\approx N$. More precisely, for the parameters given in the caption of Fig. \ref{fig:SFF}, the value of SFF at the onset of the ramp is approximately $10^{-1300}$ and rises to nearly $10^{-1260}$ where the plateau starts. As shown in the inset of the right panel, the harmonic lattice model exhibits no indication of a ramp during the SFF evolution.

%%%%%%%%%%%%%%%%%%%%%%%%%%%%%%%%%%%%%%%%%%%%%%%%%%%%%%%%%%%%%%%%%%%%%%%%%%%%%%%%%%%%%%%%%%%%%%%%%%%%%%%
\section{Information scrambling in terms of the R\'enyi-2 entropy}
\label{sec:renyi-2}
We observe that replacing the von Neumann entropy by R\'enyi-2 entropy leads to similar results for information scrambling in Gaussian states. The R\'enyi-2 entropy for Gaussian states takes a simple form in terms of the determinant of the covariance matrix of the state~\cite{Adesso2012}
\begin{equation}
    S^{(2)}(\rho)=-\ln \tr(\rho^2)= \frac12 \ln \det(\bm\sigma).
\end{equation}
As shown in Fig.~\ref{fig:Renyi2}, the dynamics of the entanglement and the TMI in terms of the R\'enyi-2 entropy have similar behavior to the corresponding quantities in terms of the von Neumann entropy. 

The Wigner function of Gaussian states is a normalized Gaussian function
\begin{equation}
    W_\rho(\bm\alpha)=\frac{1}{\pi^N \sqrt{\det(\bm\sigma)}} e^{-\bm\alpha^\top \bm\sigma^{-1} \bm\alpha},          
\end{equation}
where $\bm\alpha\in \mathbb{R}^{2N}$. Hence, the Gaussian Wigner function can be viewed as a probability density, and its continuous Shannon entropy up to an additional constant is equal to the R\'enyi-2 entropy of the state~\cite{Adesso2012},
\begin{align}\label{eq:cont-Shan}
    \begin{split}
    H(W_\rho)&=-\int d^{2N}\bm\alpha\; W_\rho(\bm\alpha) \ln \!\big(W_\rho(\bm\alpha) \big)
    \\&
    = S^{(2)}(\rho) + N(1+ \ln\pi).
    \end{split}
\end{align}
This is in fact the entropy of the classical random variable $\bm\alpha$ sampled from the Wigner function. 

\begin{figure}
\begin{center}
\includegraphics[scale=0.6]{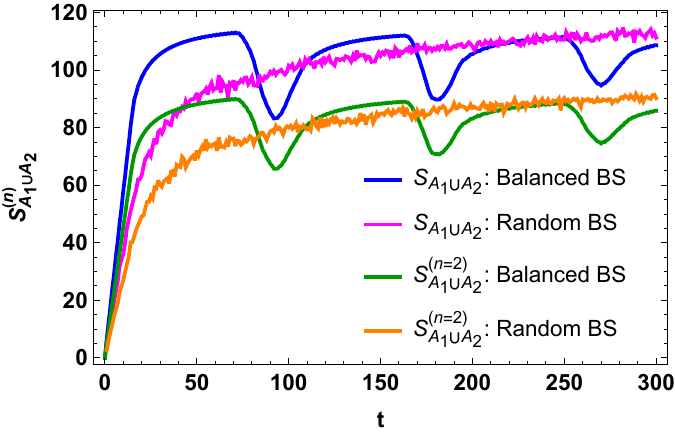}
\includegraphics[scale=0.4]{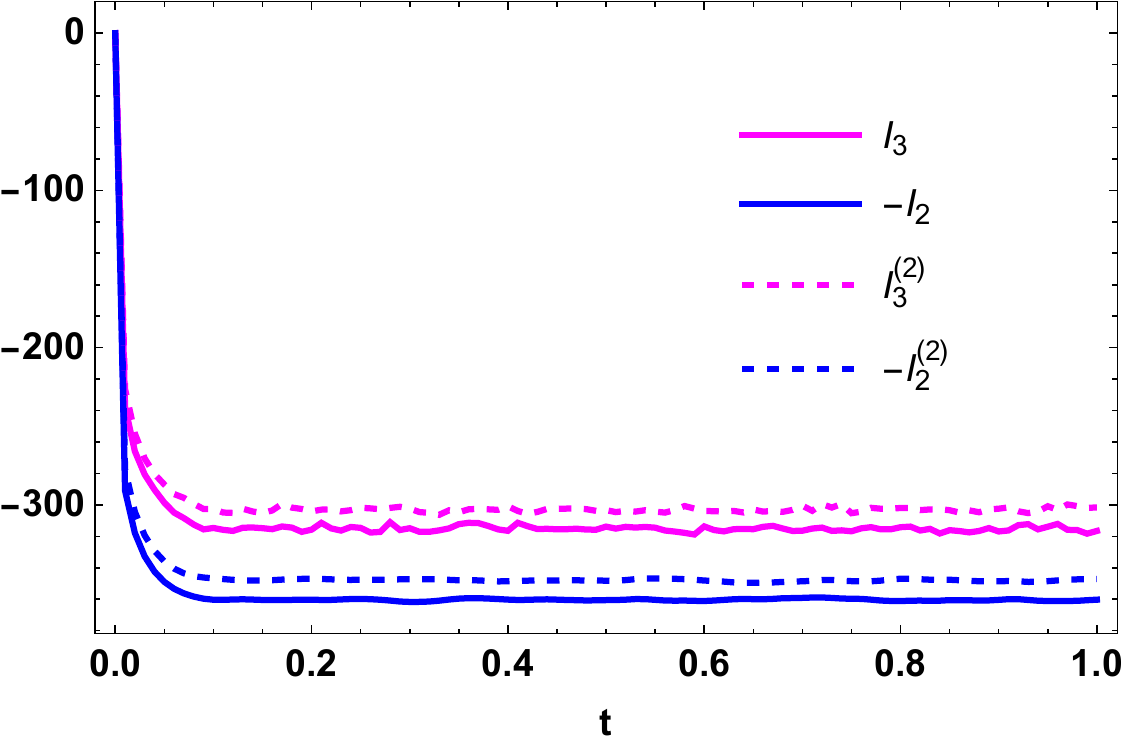}
\end{center}
\caption{Upper panel: Memory effect in the passive circuit model (IIa) compared with R\'enyi-2. The initial state is the tensor product of single-mode squeezed vacuum states with $\lambda_i=2$. We set $N_{A_1}=N_{A_2}=40$ and $d=200$. We consider the BS transfer matrix $\bm U_{\mathrm{BS}}$ chosen randomly at every single time step. For a balanced BS, we observe memory effects for both the entanglement and the R\'enyi-2 case. The second and third dips in the blue and green curves correspond to entanglement revivals. All random results are averaged over 300 samples. Lower panel: R\'enyi-2 version of TMI $I_3^{(2)}$ compared with TMI.}
\label{fig:Renyi2}
\end{figure}

This relation between the continuous Shannon entropy and the R\'enyi-2 entropy implies that information scrambling in our quantum setup can analogously be observed in a classical setup involving Gaussian random variables. Specifically, one can generate Gaussian random variables according to the Wigner function of the initial state, which are then evolved according to the Gaussian dynamics. Given Eq.~(\ref{eq:cont-Shan}) and the results presented in Fig.~\ref{fig:Renyi2}, it can be seen that similar information scrambling characteristics can be observed in terms of the entropy of the Gaussian random variables in analogous classical setups.  

%%%%%%%%%%%%%%%%%%%%%%%%%%%%%%%%%%%%%%%%%%%%%%%%%%%%%%%%%%%%%%%%%%%%%%%%%%%%%%%%%%%%%%%%%%%%%%%%%%%%%%%
\section{Discussions}
\label{sec:disc}
We have shown that randomness in integrable bosonic systems leads to information scrambling, characterized by the disappearance of the memory effect and large negative values of TMI during entanglement dynamics. These results identify information scrambling associated with randomness as a distinct feature from quantum chaos. However, we show that random quadratic Hamiltonians can have a chaotic component, causing the system to behave chaotically for certain non-Gaussian states. 

Our results reveal new aspects of the dynamics of Gaussian states in continuous-variable systems, which are useful for quantum information processing due to their experimental accessibility.  
As we have shown, similar results can be obtained in terms of the R\'enyi-2 entropy, which is linked to the continuous Shannon entropy of classical random variables generated according to the Wigner function~\cite{Adesso2012}. This implies that information scrambling characteristics can also be observed in classical systems.

It is also worth noting that the authors of \cite{Alba:2019ybw} introduced the concept of \textit{weak scrambling}, defined by the disappearance of the MI peak at infinite subregion separation in local integrable systems. This effect is further amplified in systems with non-linear dispersion relations \cite{Mozaffar:2021nex}. It is important to emphasize that our study, which focuses on the role of randomness, should not be conflated with this inherently \textit{nonrandom} phenomenon.

\section*{Acknowledgments} We thank Reza Mohammadi-Mozaffar, Salman Beigi and Pratik Nandy for useful discussions. AM would like to acknowledge support from ICTP through the Associates Programme (2023-2028) and for hospitality during stages of this work. 
  
\bibliography{bibl}

\clearpage

\appendix

\onecolumngrid

\section{More on Quadratic Hamiltonians and Gaussian States}
\subsection{Preliminaries}
We consider $N$-mode bosonic systems described by the vector of quadrature operators $r=(q_1,\dots,q_N,p_1,\dots,p_N)^\top$, satisfying the canonical commutation relations $[q_j,p_k]=i \delta_{jk}$ ($\hbar=1$). Quantum states are described by density operators $\rho$  ($\rho\succeq 0$ and $\tr(\rho)=1$), and those that can be represented by a Gaussian Wigner function are known as Gaussian states~\cite{WeedbrookRMP2012,Serafini17}. These states can be uniquely characterized using the first-order moment vector $\langle r\rangle=\tr(\rho r)$ and the covariance matrix with the matrix elements given by $\bm\sigma_{jk}=\la r_j r_k+r_k r_j\rangle-2\la r_j\lb \la r_k\lb$. The covariance matrix is symmetric and positive and satisfies the uncertainty relation $\bm\sigma+i \bm{J}\geq0$, where $\bm J$ is the symplectic form whose matrix elements are given by $\bm J_{jk}=i[r_k,r_j]$. Reduced states of a Gaussian state with the covariance matrix $\bm\sigma$ are Gaussian states whose covariance matrices are submatrices of $\bm\sigma$. We consider states with $\langle r\rangle=0$, as this condition can always be satisfied by using local unitary displacement operations. If a covariance matrix satisfies $\bm\sigma-\mathbb{1}_{2N}\geq0$, where $\mathbb{1}_{2N}$ is the $2N\times 2N$ identity matrix, the corresponding state is {\it classical} and can be expressed as a statistical mixture of product coherent states. A special example is the product thermal state of $N$-mode system whose $\langle r\rangle=0$ and $\bm\sigma_{\rm{th}}=\text{diag}(\nu_1,\dots, \nu_N,\nu_1,\dots, \nu_N)$, where $\nu_k \geq 1$ with $\nu_k=1$ corresponding to the vacuum state in the $k$th mode.

Unitary transformations that preserve the Gaussianity of quantum states are known as Gaussian unitaries. Up to local displacement operations, Gaussian unitaries are associated with quadratic Hamiltonians of this form
\begin{equation}\label{eq:H}
H=\frac{1}{2}r^\top \bm M r
\end{equation}
with matrix $\bm M$ being real, symmetric, and positive definite~\footnote{In general, $\bm M$ can be a Hermitian matrix, which can then be written as a sum of real symmetric and anti-symmetric matrices, $\bm M=\bm M_{\rm{sym}}+i \bm M_{\rm{asy}}$. However, the anti-symmetric part $\bm M_{\rm{asy}}$ only contributes an overall phase that can be ignored.}. Such Gaussian unitary can be described by $e^{iHt} r e^{-iHt}=\bm S r$, where  $ \bm S=e^{\bm J\bm Mt} $ is a symplectic matrix, satisfying $\bm S \bm J \bm S^\top =\bm J$. Using this, one can see that the evolution of Gaussian states under Gaussian unitaries, $\rho_t =e^{-iHt} \rho e^{iHt}$, can be described in terms of symplectic transformations on the covariance matrices, $\bm \sigma_t = \bm S \bm\sigma \bm S^\top$. Gaussian unitaries with orthogonal symplectic matrices, preserve the mean energy of the system and are known as passive transformations. In quantum optics, these transformations %, known as linear-optical networks, 
can be realized using linear-optical networks.% lossless beam splitters and phase rotations. 

According to the Euler decomposition (see for instance \cite{Serafini17}), any Gaussian unitary can be expressed as $\smash{U_{\rm p}\otimes_{j=1}^{N} U_{\text{sq},\lambda_j} \tilde U_{\rm p}}$, where $U_{\rm p}$ and $\tilde U_{\rm p}$ are passive multimode unitaries, and $U_{\text{sq},\lambda_j}$ are single-mode squeezing unitaries, described by $U_{\text{sq},\lambda_j}^\dagger (q_j,p_j) U_{\text{sq},\lambda_j}=(q_j e^{-\lambda_j},p_je^{\lambda_j})$. Williamson’s theorem \cite{Williamson} states that a Gaussian unitary can transform any Gaussian state into a thermal state. This implies that covariance matrices can be diagonalized using symplectic transformations, $\bm \sigma=\bm S\, \text{diag}(\nu_1,\dots, \nu_N,\nu_1,\dots, \nu_N) \bm S^\top$, where $\nu_k$ are known as the symplectic eigenvalues. Considering this theorem and noting that passive unitaries do not change the vacuum state, we can see that any pure Gaussian state can be constructed by applying single-mode squeezing unitaries on the vacuum state followed by a multimode passive Gaussian unitary. Note also that one can see that the von Neumann entropy of Gaussian states can be expressed in terms of the symplectic eigenvalues
\begin{equation}
S_{\text{vN}}=
    \sum_k \bigg[\frac{\nu_k+1}{2} \ln\! \left(\frac{\nu_k+1}{2}\right)- \frac{\nu_k-1}{2} \ln\!\left(\frac{\nu_k-1}{2}\right)\!\bigg]\,.
\end{equation}

Gaussian states can also be viewed as a thermal state of a quadratic Hamiltonian $\rho=e^{-\beta H}/\tr(e^{-\beta H})$~\cite{Serafini17}. In this view, Williamson’s theorem implies that the Hamiltonian can be uncoupled by a Gaussian unitary $U^\dagger H U= \sum_{j=1}^N \omega_j(q_j^2+p_j^2)$, where $\omega_j$ are the frequencies of the uncoupled modes, known as normal modes. Here, the Gaussian unitary $U$ corresponds to the symplectic transformation that diagonalizes the Hamiltonian matrix $\bm S^\top\bm M\bm S=\text{diag}(\omega_1,\dots \omega_N,\omega_1,\dots \omega_N)$. Using this Gaussian unitary, we can obtain the Hamiltonian's ground state from the uncoupled Hamiltonian's vacuum state.

\subsection{Harmonic lattice model: symplectic transformation and the ground state}
The Hamiltonian matrix  $\bm M_{qp}=0$, $\bm M_{p}=\mathbb{1}_N/\epsilon$, and $\bm M_{q}=1/\epsilon\; \mathrm{circ}(\epsilon^2 m^2+2,-1,\cdots,-1)$, where $\mathrm{circ}$ demotes the circulant matrix. The eigenvalues of the HL model are $\smash{m^{2}+\big(2\sin({\pi (j-1)}/{N})\big)^{2}}$ for $j=1,\dots,N$. By ordering the eigenvalues $\smash{\omega_1^2\,{\leq}\,\omega_2^2\,{\leq}\cdots{\leq}\,\omega_N^2}$, an orthogonal matrix $\bm V$ that diagonalizes this matrix can be found as follows. Defining row vectors $\smash{F_j=\frac{1}{\sqrt{N}}(1,\nu^{(N+1-j)}, \dots,\nu^{(N-1)(N+1-j)})}$ with $\nu=\exp(2\pi i/N)$, the first row of $\bm V$ is given by $V_1=F_1$, other rows are given by $V_{2k}=({F_{k+1}+F_{N-k+1}})/\sqrt{2}$ and $V_{2k+1}=i({F_{k+1}-F_{N-k+1}})/{\sqrt{2}}$ for $1\leq k< N/2$, and if $N$ is even the last row becomes $V_N=F_{\frac{N}{2}+1}$. Using this, the HL Hamiltonian matrix  can be diagonalized by the symplectic matrix $\bm S_{\rm{p}}=\bm V\oplus \bm V$, corresponding to a passive Gaussian unitary $U_{\rm p}$. Applying this unitary on the Hamiltonian gives $U_{\rm p}^\dagger H U_{\rm p}=\sum_{j=1}^N (\omega_j^2q_j^2+p_j^2)$. Therefore, by using additional local squeezing transformations $\otimes_{j=1}^{N} U_{\text{sq},\lambda_j}$ with squeezing parameters $\lambda_j=\frac{1}{2}\log\omega_j$, we obtain the Hamiltonian of uncoupled harmonic oscillators $\sum_{j=1}^N \omega_j^2(q_j^2+p_j^2)$.

The ground state of this Hamiltonian is a Gaussian state of $N$ mode, which can be prepared in quantum optic settings by applying local squeezing operations $\otimes_{j=1}^{N} U_{\text{sq},\lambda_j}$ on the vacuum state of each mode, followed by an $N$-mode passive linear optical transformation described by orthogonal matrix $\bm V$. This analogy provides an operational cost of reaching the conformal field theory regime. Note that to reach this scale-invariant regime, we need to take the limit of $m\to 0$. However, since $\omega_1=m$, taking this limit implies infinite squeezing $\lambda_1\to-\infty$, meaning that preparing the ground state in the scale-invariant regime requires infinite energy.

\section{Tripartite Mutual Information}

\subsection{Evolution under active unitary transformations}

In the main text, we discussed the time evolution of TMI under passive unitary transformations, where a well-defined Haar measure imposes a lower bound on TMI. These passive transformations correspond to our second family of models (IIa, IIb). By contrast, the first family of models (Ia, Ib), which may be more familiar to readers with backgrounds in many-body physics and high-energy theory, correspond to active transformations and active unitary time evolutions. Unlike passive transformations, which conserve the number of particles in the state, active transformations do not. Additionally in contrast to passive transformations, the symplectic transformation corresponding to active transformations is not orthogonal.

\begin{figure}[h]
\begin{center}
\includegraphics[scale=0.48]{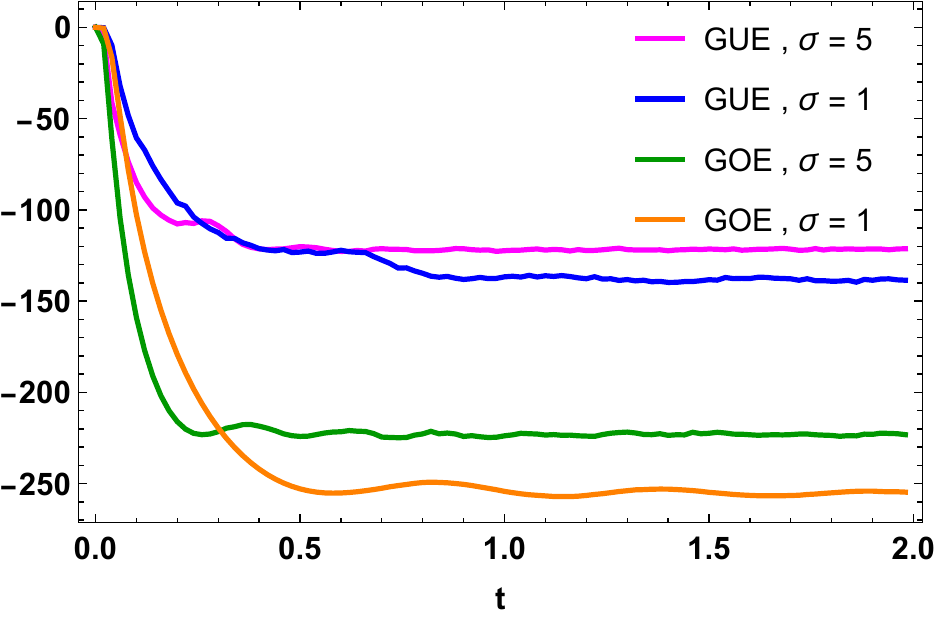}
\hspace{5mm}
\includegraphics[scale=0.48]{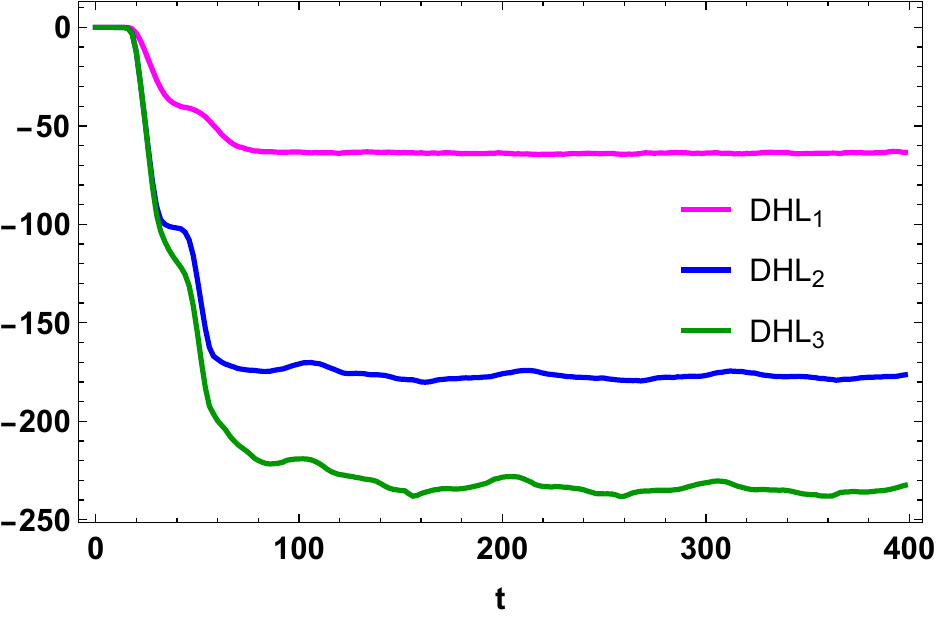}
\end{center}
\caption{
TMI active random unitary time evolution. The initial state is a uniform squeezed state with $\lambda=10$ and we set $N=200$ with $N_A=20$, $N_B=80$, and $N_C=30$. Left: Nonlocal random dynamics driven by (Ib) with GUE and GOE, and varying the variance $\sigma$. The different saturation values demonstrate that there is no lower bound for TMI. Right: Local random dynamics driven by DHL (Ia) where DHL$_1$ corresponds to $J\in(0,2)$, DHL$_2$ corresponds to $J\in(0.5,1.5)$, and DHL$_3$ corresponds to $J\in(0.75,1.25)$. In all plots are averaged over 100 samples.}
\label{fig:I3active}
\end{figure}

Here, we show that the behavior of TMI under unitary time evolutions governed by the disordered harmonic lattice model (Ia) and random non-local Hamiltonians (Ib) shares similar features with passive dynamics: starting from an adequately entangled state, under (Ib) dynamics, TMI rapidly saturates to a negative value, and the same occurs for (Ia) on a larger timescale. The important difference is that for non-passive dynamics, there is no lower bound for TMI, reflecting the fact that a Haar measure is not well-defined for this set of unitaries (see Fig. \ref{fig:I3active}).

\subsection{TMI for static states}
In the main text, we examined the behavior of TMI under unitary dynamics for a fixed ratio between subregions. Here, we briefly discuss the sign of TMI for configurations with varying subregion ratios in a static state. As illustrated in Fig. \ref{fig:Renyi2}, we demonstrate that, irrespective of the ratio, the sign of TMI evolved by a Haar unitary remains consistently negative. Interestingly, as shown by the dashed curves in the left panel, even after introducing shot noise—which disrupts all quantum correlations—we still observe negative TMI values, albeit with significantly reduced magnitudes. Furthermore, as indicated by the dashed curves in the right panel, the R\'enyi-2 variant of TMI exhibits a similar behavior, suggesting a possible interpretation in terms of classical random Gaussian variables.

\begin{figure}
\begin{center}
\includegraphics[scale=0.65]{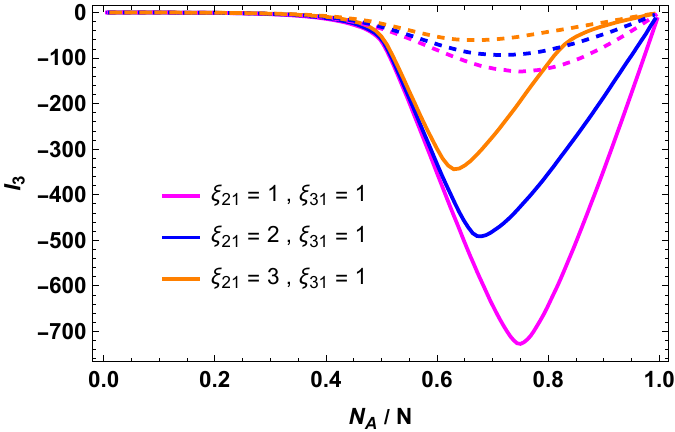}
\hspace{5mm}
\includegraphics[scale=0.65]{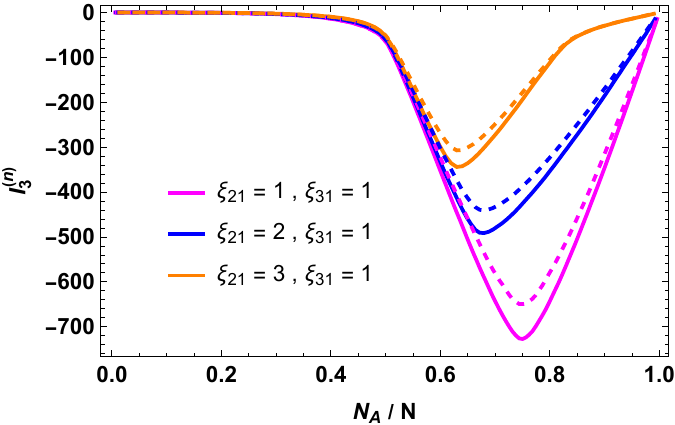}
\end{center}
\caption{Left: Entanglement in the passive circuit model (IIb) for disjoint blocks compared with R\'enyi-2. The initial state is the tensor product of single-mode squeezed vacuum states with $\lambda_i=2$. We set $N_{A_1}=N_{A_2}=40$ and $d=200$. We consider the BS transfer matrix $\bm U_{\mathrm{BS}}$ chosen randomly at every single time step. For a balanced BS, we observe memory effects for both the entanglement and the R\'enyi-2 case. The second and third dips in the blue and green curves correspond to entanglement revivals. All random results are averaged over 300 samples. Right: $I_3$ for random model (IIa), where the tensor product of squeezed-vacuum states ($\lambda_i=5$) is evolved by a one-step $N$-mode passive network described by a Haar random unitary. We set $N=500$ and $\xi_{ij}\equiv\frac{N_{A_i}}{N_{A_j}}$. The dashed curves correspond to the R\'enyi-2 version of TMI with the same parameters as the solid curves.
Lower panel: $I_3$ for random modell (IIa), where the tensor product of squeezed-vacuum states ($\lambda_i=5$) is evolved by a one-step $N$-mode passive network described by a Haar random unitary. We set $N=500$ and $\xi_{ij}\equiv\frac{N_{A_i}}{N_{A_j}}$. The dashed curves correspond to the same parameters as the solid curves but one unit of vacuum noise is added to the initial squeezed-vacuum states to make them classical. In this case, no entanglement is generated in the network.}
\label{fig:Renyi2}
\end{figure}

\section{Out of Time-Ordered Correlators}\label{sec:AppOTOC}

\subsection{Derivation of OTOC for quadrature operators}

We derive Eq.~\eqref{eq:C1}, OTOC for canonical operators, and the ground state of quadratic Hamiltonians whose matrix $\bm M$ can be diagonalized using symplectic transformations of the form $\bm S=\bm S_q\oplus \bm S_p$. This includes KG, DKG, and our random model (IIIb) Hamiltonians. We define
\begin{equation}
H_{\rm D}=U^\dagger H U=\frac{1}{2} U^\dagger r^\top U \bm M U^\dagger r U=\frac{1}{2}r^\top \bm D r
\end{equation}
where we used $U^\dagger r U= \bm S r$ and matrix $\bm D=\bm S^\top \bm M \bm S$ is diagonal.  

Assuming that $\bm S=\bm S_q\oplus \bm S_p$, we have 
\begin{align}
\Big[e^{i H_{\rm D} t} U^\dagger q_j U e^{-i H_{\rm D} t}, U^\dagger p_k U \Big]
&=\sum_{n,m}\bm S_{q,jn}\bm S_{p,km} \Big[e^{i H_{\rm D} t} q_n e^{-i H_{\rm D} t},p_m\Big]\nonumber
\\
&=\sum_{n,m} \bm S_{q,jn}\bm S_{p,km} \big[q_n\cos(\omega_k t) + p_n\sin(\omega_k t),p_m\big]\nonumber
\\
&=i \sum_{n}\bm S_{q,jn}\bm S_{p,kn}\cos(\omega_n t)\nonumber
\\
&=i \Big(\bm S_q\, \text{diag}\big(\cos(\omega_1 t), \cdots,\cos(\omega_N t)\big) \bm S_p^\top \Big)_{ij},\nonumber
\end{align}
where we used $e^{i H_{\rm D} t} q_n e^{-i H_{\rm D}t}=q_n\cos(\omega_k t) + p_n\sin(\omega_k t)$ in the second line, and $[q_n,p_m]=i\delta_{nm}$ in the third line. Using the above relation and $H=U H_{\rm D} U^\dagger$, OTOC for canonical operators reads
\begin{align}
    C_{\beta,jk}(t)&=-\tr\!\left(\frac{e^{-\beta H}}{\tr\left(e^{-\beta H}\right)}\big[q_j(t),p_k\big]^2\right)\nonumber \\
    &= -\tr \left(\frac{e^{-\beta H_{\rm D}}}{\tr\left(e^{-\beta H_{\rm D}}\right)} \Big[e^{i H_{\rm D} t} U^\dagger q_j U e^{-i H_{\rm D} t},U^\dagger p_j U\Big]^2\right)\nonumber \\
    &= \Big(\bm S_q\, \text{diag}\big(\cos(\omega_1 t), \cdots,\cos(\omega_N t)\big) \bm S_p^\top\Big)_{ij}^2,
\end{align}
which is Eq.~\eqref{eq:C1} in the main text. This expression can be viewed as a generalization of the same quantity for a single oscillator previously reported in \cite{Hashimoto:2017oit}.

%%%%%%%%%%%%%%%%%%%%%%%%%%%%%%%%%%%%%%%%%%%%%
%%%%%%%%%%%%%%%%%%%%%%%%%%%%%%%%%%%%%%%%%%%%%
\subsection{OTOC for generic bosonic operators under passive dynamics}
Considering an $N$-mode bosonic system with the Hamiltonian $H$ and local operators $W_j$ and $V_k$ acting on the $j$th and $k$th modes, respectively, OTOC can alternatively be defined as
\begin{equation}\label{eq:f-otoc}
    F_{\beta,jk}(t)=\left\langle W^\dagger_j(t) V^\dagger_k W_j(t) V_k \right\rangle=\tr \Bigg(\! \frac{e^{-\beta H}}{\tr(e^{-\beta H})} W^\dagger_j(t) V^\dagger_k W_j(t) V_k \!\Bigg),
\end{equation}
where $W_j(t)=e^{-it H} W_j e^{it H}$. Note that if $W_j$ and $V_k$ are unitary, then the relation between this quantity and $C_{\beta,jk}(t)=\big\langle [V_k,  W_j(t)]^\dagger [V_k,  W_j(t)] \big\rangle$, which is used in the main text, is given by $\text{Re}\big[F_{\beta,jk}(t)\big]=1-C_{\beta,jk}(t)/2$. Here, we consider passive Hamiltonians that can be transformed into an uncoupled Hamiltonian of harmonic oscillators using a passive Gaussian unitary,
\begin{equation}\label{eq:pass-hamiltonian}
    H=(a_1,\dots,a_N)\bm M (a_1^\dagger,\dots,a_N^\dagger)^\top=U_\text{p} H_\text{D} U_\text{p}^\dagger
\end{equation}
with $H_\text D=\sum_j \omega_j a^\dagger_j a_j$. The passive unitary can be described by $U_\text p^\dagger a^\dagger_j U_\text p=\sum_n  \bm U_{jn} a^\dagger_n $ with $\bm U_{jn}$ being the elements of the $N\times N$ unitary matrix that diagonalizes the Hamiltonian matrix, $\sum_{jk}\bm U^*_{mk} \bm M_{kj} \bm U_{jn}=\omega_n\delta_{nm}$. By using this transformation, we have
\begin{equation}
    F_{\beta,jk}(t)=\tr \Bigg(\! \frac{e^{-\beta H_\text D}}{\tr(e^{-\beta H_\text D})} \tilde W^\dagger_j(t) \tilde V^\dagger_k \tilde W_j(t) \tilde V_k \! \Bigg)
\end{equation}
where $\tilde W_j(t)=e^{it H_\text D} U_\text p^\dagger W_j U_\text p e^{-it H_\text D}$ and $\tilde V_k=U_\text p^\dagger V_k U_\text p$.

The thermal state of $N$-mode uncoupled system can be expressed in terms of multimode coherent states $\ket{\alpha_1,\dots,\alpha_N}$ using the Glauber-Sudarshan representation~\cite{Glauber1963,Sudarshan1963}
\begin{equation}\label{eq:thermal-glauber-sudarshan}
    \frac{e^{-\beta H_\text D}}{\tr(e^{-\beta H_\text D})}=\idotsint d^2\alpha_1 \dots d^2\alpha_N \frac{e^{-|\alpha_1|^2/\bar{n}_1}}{\bar{n}_1 \pi}\times \dots \times \frac{e^{-|\alpha_N|^2/\bar{n}_N}}{\bar{n}_N \pi} \ket{\alpha_1,\dots,\alpha_N}\bra{\alpha_1,\dots,\alpha_N}
\end{equation}
with $\bar{n}_j=(e^{\beta\omega_j}-1)^{-1}$ being the mean photon number in the $j$th mode. Using this, we can write
\begin{equation}\label{eq:otoc-f-alpha}
        F_{\beta,jk}(t)=\int d^{2N}\!\bm\alpha\; P(\bm\alpha) \tilde F_{\bm\alpha,jk}(t),
\end{equation}
where $\bm\alpha=(\alpha_1,\dots,\alpha_N)$, $P(\bm\alpha)=\prod_{j=1}^N \exp\!\big(\!-|\alpha_j|^2/\bar{n}_j\big)/(\bar{n}_j \pi)$, and 
\begin{equation}\label{eq:coherent-otoc}
    \tilde F_{\bm\alpha,jk}(t)= \big\langle\alpha_1,\dots,\alpha_N\big| \tilde W^\dagger_j(t) \tilde V^\dagger_k \tilde W_j(t) \tilde V_k \big|\alpha_1,\dots,\alpha_N\big\rangle
\end{equation}
is the OTOC in terms of coherent states. Note that for $\beta=\infty$ (zero temperature) $F_{\infty,jk}(t)=\tilde F_{\bm 0,jk}(t)$.

The operators can also be expanded in terms of displacement operators $D_j(\zeta)=\exp(\zeta a^\dagger_j -\zeta^* a_j)$ with $\zeta\in \mathbb{C}$ \cite{Cahill1969}
\begin{equation}
    W_j=\frac{1}{\pi}\int d^2\zeta \; \Phi_{W_j}\!(\zeta)\; D_j(-\zeta),
\end{equation}
where $\Phi_{W_j}\!(\zeta)=\tr \big(W_j D_j(\zeta)\big)$ is the characteristic function of operator $W_j$. We have 
\begin{equation}
    U_\text p^\dagger D_j(\zeta) U_\text p=D_1(\zeta \bm U_{j1})\otimes \dots \otimes D_N(\zeta \bm U_{jN}),
\end{equation}
where we used $U_\text p^\dagger a^\dagger_j U_\text p=\sum_n  \bm U_{jn} a^\dagger_n $, and also
\begin{equation}
    e^{it H_\text D} D_j(\zeta) e^{-it H_\text D}=D_j(\zeta e^{it\omega_j}).
\end{equation}
Using these two relations we can then compute
\begin{equation}
    \tilde W_j(t)=e^{it H_\text D} U_\text p^\dagger W_j U_\text p e^{-it H_\text D}=\frac{1}{\pi} \int d^2\zeta \; \Phi_{W_j}\!(\zeta) D_1(-\zeta \bm U_{j1} e^{it\omega_1})\otimes \dots \otimes D_N(-\zeta \bm U_{jN} e^{it\omega_N}).
\end{equation}
Therefore, by using a similar expression for $\tilde V_k$, the OTOC in terms of coherent states can be written as
\begin{equation}\label{eq:coherent-otoc-charac}
\begin{split}
      \tilde F_{\bm\alpha,jk}(t)=\frac{1}{\pi^4} &\int d^2\zeta d^2\xi d^2\beta d^2\gamma\; \Phi^*_{W_j}\!(\zeta) \Phi^*_{V_k}\!(\xi)
\Phi_{W_j}\!(\beta) \Phi_{V_k}\!(\gamma) \\
&\times\prod_{n=1}^N \big\langle\alpha_n\big| D_n(\zeta \bm U_{jn} e^{it\omega_n}) D_n(\xi \bm U_{kn})  D_n(-\beta \bm U_{jn} e^{it\omega_n}) D_n(-\gamma \bm U_{kn}) \big|\alpha_n\big\rangle.
\end{split}
  \end{equation}
As we can see each term in the product oscillates with time with frequency $\omega_n$. By using the following relation for displacement operators 
\begin{equation}
    D_n(\mu) D_n(\nu)= \exp\!\bigg(\! \frac{\mu \nu^*-\nu\mu^*}{2} \!\bigg) D_n(\mu+\nu), 
\end{equation}
we get
\begin{equation}
    \begin{split}
        D_n(\zeta \bm U_{jn} &e^{it\omega_n}) D_n(\xi \bm U_{kn})  D_n(-\beta \bm U_{jn} e^{it\omega_n}) D_n(-\gamma \bm U_{kn}) = D_n\big( \bm U_{jn} e^{it\omega_n} (\zeta-\beta) + \bm U_{kn} (\xi-\gamma)\big)  \\
        &\times \exp\bigg[\frac12 \Big((\zeta\xi^*+\beta\gamma^*+\beta\xi^*-\zeta\gamma^*)\bm U_{jn}e^{i\omega_n t}\bm U_{kn}^*+\gamma\xi^*|\bm U_{kn}|^2+ \beta\zeta^*|\bm U_{jn}|^2\Big)-\text{c.c.}\bigg], 
        \end{split}
\end{equation}
where c.c. stands for complex conjugate. We also have $\bra{\alpha}D(\mu)\ket{\alpha}=\exp(-|\mu|^2/2+\mu\alpha^*-\alpha\mu^*)$. Using these equations, Eq.~(\ref{eq:coherent-otoc-charac}) becomes
\begin{align}
\begin{split}\label{eq:FOTOC}
   \tilde F_{\bm\alpha,jk}(t)= \frac{1}{\pi^4}&\int d^2\zeta d^2\xi d^2\beta d^2\gamma\; \Phi^*_{W_j}\!(\zeta) \Phi^*_{V_k}\!(\xi)
\Phi_{W_j}\!(\beta) \Phi_{V_k}\!(\gamma)\\
&\times\exp\!\bigg[\!-\frac12 |\zeta-\beta|^2-\frac12 |\xi-\gamma|^2 -\frac12 \big((\zeta-\beta)(\xi^*-\gamma^*) \bm V_{jk}(t) +\text{c.c.} \big)\bigg]
%\nonumber
\\
&\times \exp\bigg[\Big((\zeta-\beta)  \sum_{n=1}^N \bm U_{jn} e^{it\omega_n} \alpha_n^* + (\xi-\gamma)  \sum_{n=1}^N \bm U_{kn} \alpha_n^* \Big)-\text{c.c.} \bigg]
%\nonumber
\\
&\times \exp\!\bigg[\frac12\Big((\zeta\xi^*+\beta\gamma^*+\beta\xi^*-\zeta\gamma^*) \bm V_{jk}(t)  +\gamma\xi^*+\beta\zeta^*\Big)-\text{c.c.}  \bigg],
%\nonumber
\end{split}
\end{align}
where we used $\sum_n|\bm U_{jn}|^2=1$ and 
\begin{equation}
    \bm V_{jk}(t)=\sum_{n=1}^N \bm U_{jn} e^{i\omega_n t} \bm U_{kn}^*
\end{equation}
is an element of the unitary matrix $\bm V(t)=\exp(i\bm Mt)$, where $\bm M$ is the Hamiltonian matrix in Eq.~(\ref{eq:pass-hamiltonian}). Given the characteristic functions $\Phi_{V_k}$ and $\Phi_{W_j}$, one can therefore compute $F_{\bm\alpha,jk}(t)$ and then by averaging with respect to $P(\bm\alpha)$ in Eq.~(\ref{eq:otoc-f-alpha}), $F_{\beta,jk}(t)$ can be obtained.  

As a simple example, suppose that $W_j=D_j(\mu)$ and $V_k=D_k(\nu)$ are displacement operators with the characteristic functions $\Phi_{W_j}(\beta)=\pi \delta^2(\beta+\mu)$ and $\Phi_{V_k}(\gamma)=\pi\delta^2(\gamma+\nu)$. In this case, we have
\begin{equation}
     \tilde F_{\bm\alpha,jk}(t)=\exp\!\big(\mu\nu^* \bm V_{jk}(t) -\mu^*\nu \bm V^*_{jk}(t) \big),
\end{equation}
which implies that $F_{\beta,jk}(t)=\tilde F_{\bm\alpha,jk}(t)$ is independent of $\beta$. This also gives $C_{\beta,jk}(t)=2-2\text{Re}[F_{\beta,jk}(t)]=4 \sin^2\!\big(\text{Im}[\mu\nu^* \bm V_{jk}(t)]\big)$. For similar expressions for OTOC corresponding to other Gaussian and non-Gaussian dynamics see~\cite{Zhuang:2019jyq}.

\begin{figure}[h]
\begin{center}
\includegraphics[scale=0.4]{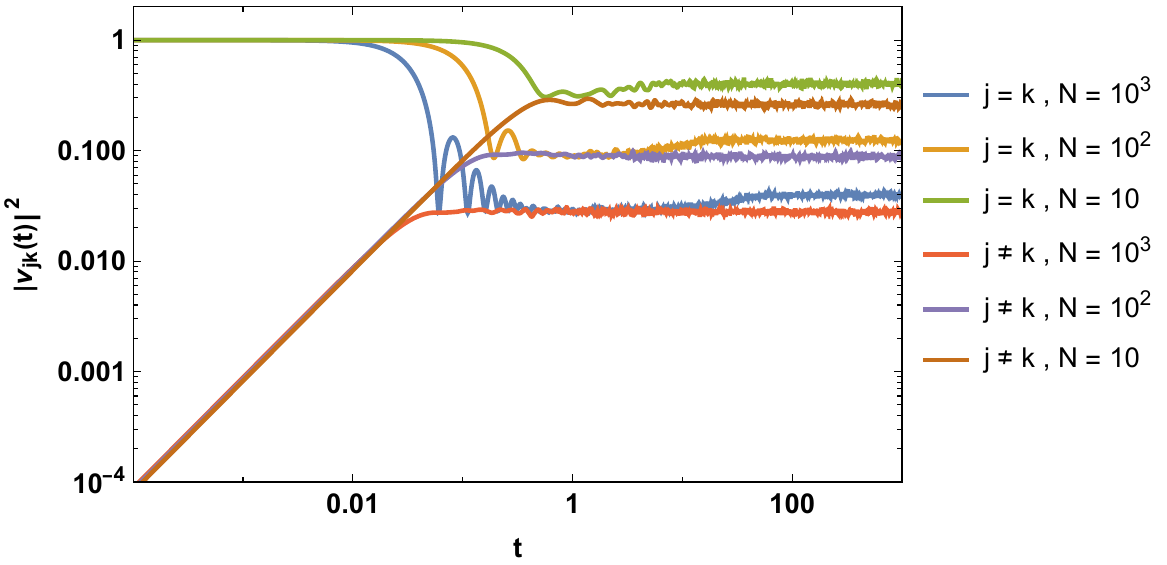}
\hspace{4mm}
\includegraphics[scale=0.38]{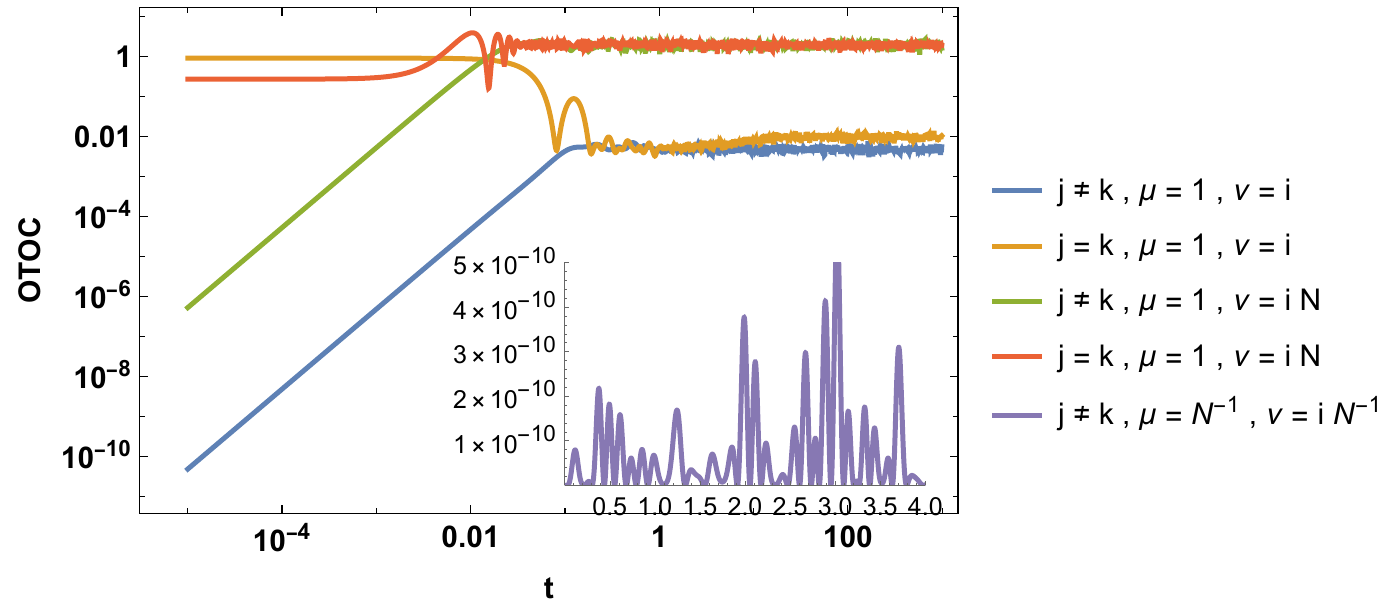}
\end{center}
\caption{Right: Time evolution of the diagonal and off-diagonal elements of $\bm V_{jk}(t)$. $\bm U$ is a Haar random unitary matrix and the result is averaged over 200 samples. Left: The behavior of OTOC $F_{\beta,jk}(t)$ for displacement operators with parameters $\mu$ and $\nu$. The inset shows the choice of $\mu\nu^*\sim 1/N^2$ is small enough to compensate for the rapid saturation of $\bm V_{jk}(t)$ and allow the OTOC to oscillate in time.} 
\label{fig:v}
\end{figure}

In this example, we can see that the time-dependent behavior of OTOC is determined by that of $\bm V_{jk}(t)$. Therefore, to further investigate the time-dependent behaviour of $\bm V_{jk}(t)$, we ran numerical simulations for a randomly selected set of $\{\omega_n\}$ and $\bm U$ selected as an $N\times N$ unitary matrix chosen from the Haar measure. As shown in the left panel of Fig.~\ref{fig:v}, $\bm V_{jk}(t)$ rapidly saturates to a constant value at some time $t_*\sim 1/\log N$. The larger the system size, the faster $\bm V_{jk}(t)$ and the OTOC saturation. This behavior is critically different from the scrambling time $t_s\sim \log N$ in generic chaotic systems. Due to this rapid saturation, OTOC does not have enough time to oscillate except for the cases where the parameters in $W_j=D_j(\mu)$ and $V_k=D_k(\nu)$ slow down this rapid saturation time (see the inset of the right panel in Fig. \ref{fig:v}).   

\begin{figure}[t]
\begin{center}
\includegraphics[scale=0.47]{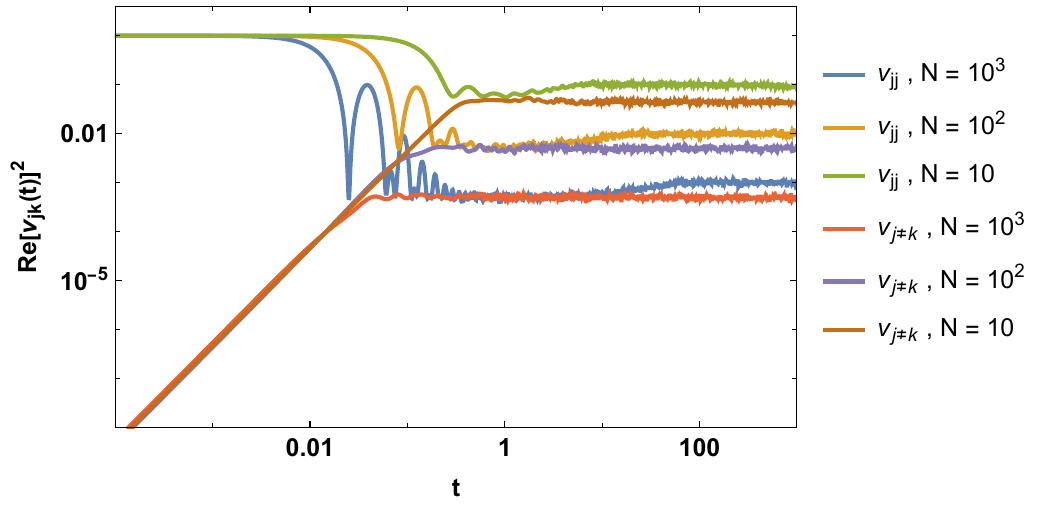}
\end{center}
\caption{The behavior of OTOC $C_{\infty,jk}(t)$ for the canonical operators $V=q_j$ and $W=p_k$. The results are averaged over 300 samples.}
\label{fig:otoc-qp}
\end{figure}

Another example is to consider canonical operators as local operators $W_j=q_j=(a_j+a_j^\dagger)/\sqrt{2}$ and $V_k=p_k=-i(a_k-a_k^\dagger)/\sqrt{2}$ whose characteristic functions are $\Phi_{W_j}(\beta)=\pi (\frac{\partial}{\partial \beta}-\frac{\partial}{\partial \beta^*}) \delta^2(\beta)/\sqrt{2}$ and $\Phi_{V_k}(\gamma)=i \pi (\frac{\partial}{\partial \gamma}+\frac{\partial}{\partial \gamma^*})   \delta^2(\gamma)/\sqrt{2}$. Plugging these into \eqref{eq:FOTOC} for $\bm\alpha=\bm 0$ and using integration by parts yields
\begin{equation}\label{eq:fotocXandP}
    \tilde F_{\mathbf{0},jk}(t)=F_{\infty,jk}(t)=\frac{1}{4}\left(1+|\bm V_{jk}(t)|^2-{\bm V^*_{jk}}^2(t)\right)
\end{equation}
which is again governed by the behaviour of $\bm V_{jk}(t)$. Using the above formalism, it is straightforward to show that the other definition of OTOC at zero temperature is given by
\begin{equation}\label{eq:otocXandP}
    C_{\infty,jk}(t)=\frac{1}{4}\left(2|\bm V_{jk}(t)|^2+\bm V_{jk}^2(t)+{\bm V^*_{jk}}^2(t)\right)=\big[\mathrm{Re}(\bm V_{jk}(t))\big]^2\,.
\end{equation}
As shown in Fig.~\ref{fig:otoc-qp}, this function shows a power law growth, similar to the behavior in Fig.~\ref{fig:v}. Note that for a single oscillator, this is a periodic function (see Eq. \eqref{eq:C1} and \cite{Hashimoto:2017oit}) while for $N>1$ the role of $\bm V_{jk}(t)$ becomes important. 

The above formalism can also be applied to compute OTOC for $q_j^{n_q}$ and $p_k^{n_p}$. In this case, the characteristic functions are given in terms of higher-order partial derivatives of the Dirac delta function \eqref{eq:FOTOC} that results in an integer power of $\mathrm{Im}(\bm V_{jk}(t))$ and $\mathrm{Re}(\bm V_{jk}(t))$ in OTOC. Note that, in general, due to the rapid saturation of $\bm V_{jk}(t)$, the result of OTOC reduces to a power law in $t$.

\end{document}